\documentclass{egpubl}
\usepackage{pg2026}

\SpecialIssuePaper         

\CGFccby

\usepackage[T1]{fontenc}
\usepackage{dfadobe}  

\usepackage{cite}  
\BibtexOrBiblatex
\electronicVersion
\PrintedOrElectronic
\ifpdf \usepackage[pdftex]{graphicx} \pdfcompresslevel=9
\else \usepackage[dvips]{graphicx} \fi

\pdfpageheight=\paperheight
\pdfpagewidth=\paperwidth

\usepackage{egweblnk}
\usepackage{booktabs}
\usepackage{amsmath}
\usepackage{float}

\renewcommand{\arraystretch}{0.92}

\makeatletter
\let\oldthebibliography\thebibliography
\let\endoldthebibliography\endthebibliography
\renewenvironment{thebibliography}[1]{%
  \oldthebibliography{#1}%
  \small
  \setlength{\itemsep}{0pt plus 0.2ex}%
  \setlength{\parskip}{0pt}%
}{\endoldthebibliography}
\makeatother

\begin{document}

\title[TileGS: Tile-Local Depth Binning for Gaussian Splatting Rasterization]%
      {TileGS: Tile-Local Depth Binning for Gaussian Splatting Rasterization}
      
\author[Tan et al.]
{
  \parbox{\textwidth}{
  \centering
    Wei Tan$^{1}$\orcid{0009-0008-7106-3339}
    \quad
    Matias Turkulainen$^{1}$\orcid{0009-0007-6931-2386}
    \quad
    Lauri Ilola$^{2}$\orcid{0009-0002-7317-9294}
    \quad
    Hamed Rezazadegan Tavakoli$^{2}$\orcid{0000-0002-9466-9148}
    \quad
    Juho Kannala$^{1}$\orcid{0000-0001-5088-4041}
    \\[0.5em]
    $^{1}$Aalto University, Finland \\
    $^{2}$Nokia Technologies, Finland
  }
}

\maketitle

\begin{abstract}
Real-time 3D Gaussian Splatting (3DGS) achieves high rendering quality, but its
standard sort-then-traverse rasterization pipeline, as instantiated by gsplat,
still traverses a globally sorted tile stream that creates long per-tile ranges
and heavy geometry-attribute traffic.
We present \emph{TileGS}, a tile-local reorganization of Gaussian splatting.
TileGS turns each long tile range into a sequence of shorter depth-local ranges,
rasterizes those ranges in front-to-back order, and applies selective repair
where coarse ordering is insufficient to match baseline compositing.
Across a 9-scene benchmark on desktop and laptop Ada GPUs, our default No-GW
(No Geometry-Write) variant delivers a mean \textbf{1.44$\times$} raster-kernel
speedup on RTX~4090
and mean end-to-end frame speedups of 1.069$\times$ on RTX~4090 and
1.094$\times$ on RTX~1000~Ada over gsplat---a widely used optimized
open-source 3DGS implementation---while matching the gsplat output up to
numerical noise ($|\Delta\mathrm{PSNR}| < 0.001~\mathrm{dB}$,
$|\Delta\mathrm{SSIM}| < 0.001$, $|\Delta\mathrm{LPIPS}| < 0.001$).
Full-suite RTX~4090 Nsight Compute profiling reveals TileGS is faster despite
lower SM throughput, lower active-warp occupancy, and higher DRAM traffic, while
total SASS thread instructions fall by \textbf{1.26$\times$}.
Source-attributed profiling confirms that geometry attributes dominate the
remaining memory pressure (85.8\% of total raster traffic and 88.6\% of excess
sectors).
Together, these counters support the interpretation that TileGS improves raster
performance by reducing effective raster traversal work, rather than by reducing
byte volume, improving coalescing, increasing occupancy, or directly reducing
measured warp divergence.
\end{abstract}

\ccsdesc[500]{Computing methodologies~Rendering}
\ccsdesc[300]{Computing methodologies~Graphics processors}
\ccsdesc[300]{Computing methodologies~Rasterization}
\printccsdesc

\smallskip
\noindent\textbf{Keywords:} 3D Gaussian Splatting, novel view synthesis, depth binning, tile-local traversal, real-time rendering

\section{Introduction}
\label{sec:intro}

3D Gaussian Splatting (3DGS)~\cite{kerbl20233d} has made
high-quality radiance-field rendering practical at real-time rates, largely
because it replaces expensive volumetric ray marching with a GPU-friendly
rasterization pipeline.
Yet the rasterizer itself remains a major performance bottleneck in raster-heavy
scenes.
The key issue is that although standard 3DGS rasterizers such as gsplat assign
work to image-space tiles, the actual raster traversal is driven by a single
globally sorted tile-depth stream.
Each tile consumes one long contiguous range from this stream, whose Gaussian
IDs often refer to widely separated locations in the scene-attribute arrays.

This creates a mismatch between tile ownership and tile-local execution.
The rasterizer is tile-based in the sense that each CUDA block owns an image
tile, but it is not fully tile-local in how each tile consumes its Gaussian
workload.
This distinction---tile-based in ownership but not tile-local in traversal---is
the key insight behind TileGS.
Consecutive lanes may fetch projected means, conic parameters, colors, and
opacities at effectively unrelated Gaussian indices.
On modern GPUs, this produces long and irregular raster traversal ranges
together with heavy geometry-attribute traffic, limiting the benefit of the
otherwise tile-parallel pipeline.

TileGS treats tile-local traversal granularity as a separate design axis for
Gaussian rasterization.
The question is not only which Gaussians are assigned to each tile, but also
how each tile consumes its assigned Gaussians during front-to-back compositing.
We refer to this organization as the rasterizer's execution structure: the
granularity and ordering of traversal, the length of each tile-local range,
how that range is broken into smaller depth-local units, and how much raster
work is needed before pixels terminate.
This notion is distinct from memory coalescing.
Memory coalescing asks whether neighboring lanes access neighboring addresses;
execution structure asks whether the raster workload is organized into traversal
units that let the kernel finish useful work in fewer elapsed cycles.
As we show in \S\ref{sec:analysis}, TileGS improves raster performance by
changing traversal granularity rather than by increasing conventional occupancy,
SM throughput, or Gaussian-index coalescing counters.

We present \emph{TileGS}, named for its tile-local reorganization of Gaussian
splatting.
Its central mechanism is tile-local depth binning: each long tile range is split
into a sequence of shorter depth-local ranges before rasterization.
TileGS assigns projected Gaussian--tile entries to coarse depth bins within each
tile, builds compact bin-offset ranges, and rasterizes the bins in
front-to-back order.
The method does not change the Gaussian representation, projected inputs, or
splat evaluation equations.
Instead, it changes the order and granularity at which the raster kernel
consumes the same Gaussian workload.
It does not reduce the static set of Gaussian--tile entries assigned to a tile;
rather, it changes their tile-local ordering and grouping, which changes the
effective raster traversal work performed before per-pixel transmittance
early termination (formalized in \S\ref{sec:analysis}).
Because coarse binning alone cannot always reproduce exact front-to-back
compositing, TileGS uses selective repair to restore baseline ordering only for
the segments where coarse local ordering is insufficient.

This design is particularly motivated by raster-heavy workloads, where the
baseline globally sorted tile-depth stream produces long tile-local ranges that
amplify irregular geometry-attribute access; TileGS targets this regime by
replacing global traversal structure with tile-local depth-binned traversal, so
the largest gains appear when the raster workload is substantial enough to
amortize the added binning and repair overhead.
Across a 9-scene benchmark on RTX~4090 and RTX~1000~Ada GPUs, TileGS delivers
consistent end-to-end speedups against the optimized gsplat baseline; the
mechanism and scope of the gain are analyzed in
\S\ref{sec:results}--\ref{sec:analysis}.

\noindent\textbf{Contributions}
\begin{itemize}
  \item We identify that standard gsplat-style rasterization is tile-based in
  ownership but still globally organized in traversal, and introduce tile-local
  depth-binned rasterization while preserving the same projected Gaussian inputs
  and splat evaluation equations.

  \item We introduce selective exact repair as an enabling mechanism for using
  coarse tile-local depth bins as an execution-order optimization while
  preserving baseline-equivalent alpha compositing up to numerical noise
  ($|\Delta\mathrm{PSNR}| < 0.001~\mathrm{dB}$, $|\Delta\mathrm{SSIM}| < 0.001$,
  $|\Delta\mathrm{LPIPS}| < 0.001$).

  \item We evaluate TileGS on two Ada GPUs, obtaining end-to-end frame
  speedups of 1.069$\times$ on RTX~4090 and 1.094$\times$ on
  RTX~1000~Ada over gsplat~\cite{ye2024gsplat}---a heavily optimized
  open-source baseline that substantially outperforms the original vanilla
  3DGS implementation---across nine scenes.

  \item We show through Nsight analysis and a direct per-pixel diagnostic that
  TileGS's speedup is best explained by reduced effective raster traversal
  work, as defined in
  \S\ref{sec:analysis}: on RTX~4090 it is faster despite lower SM throughput,
  lower active-warp occupancy, and higher DRAM traffic, while total SASS thread
  instructions fall by \textbf{1.26$\times$}, corroborated by 4.77--6.20\%
  fewer Gaussian tests per pixel before termination
  (Table~\ref{tab:traversal_work}).
\end{itemize}

\noindent\textbf{Scope note.}
The current TileGS implementation targets the forward rasterization pass only.
A training-compatible version would require the backward pass to follow the same
binned and repaired ordering used in the forward pass; we leave this integration
to future work and discuss it further in \S\ref{sec:discussion}.

\section{Related work}
\label{sec:related}

TileGS targets the forward rasterization pipeline: it preserves the same
projected Gaussian inputs and compositing semantics, but replaces a globally
sorted tile-depth stream with tile-local depth binning and binned raster
traversal.
Many recent works optimize \emph{what} is rendered through pruning, compression,
or anti-aliasing, whereas TileGS optimizes \emph{how} the surviving Gaussians
are executed during rasterization.

\subsection{3D Gaussian splatting and rasterization}

3D Gaussian Splatting (3DGS)~\cite{kerbl20233d} represents a scene as a set of
anisotropic Gaussians with learned opacity, color, and covariance parameters,
and renders novel views by projecting these Gaussians to screen space and
alpha-compositing them in depth order.
Its practical significance comes from replacing expensive volumetric ray
marching with a raster-style pipeline that exposes substantial parallelism while
retaining high image quality.
The per-Gaussian projection and footprint evaluation builds on classical
point-based rendering, including Surface Splatting~\cite{zwicker2001surface}
and EWA Splatting~\cite{zwicker2002ewa}. 
TileGS leaves this footprint evaluation unchanged and operates downstream,
reorganizing the traversal order of already-projected, tile-assigned Gaussians.

Recent high-performance implementations preserve this overall rendering model
while improving the engineering of projection, culling, and rasterization.
In particular, gsplat~\cite{ye2024gsplat} provides a widely used optimized
implementation of Gaussian splatting and serves as a strong baseline for modern
3DGS rendering systems.
Its forward path first projects Gaussians, determines their tile intersections,
sorts these intersections by tile and depth, and then rasterizes a single
globally sorted tile-depth stream.
This execution model is efficient in aggregate, but it still leaves the raster
kernel with highly irregular per-warp geometry-attribute access.

\subsection{Efficient Gaussian rendering}

A complementary line of work improves Gaussian rendering by changing the
representation or the rendering formulation rather than the execution structure
of the raster pipeline.
Examples include alternative Gaussian formulations, sort-free compositing, and
geometry-aware Gaussian variants~\cite{kerbl20233d,hou2025sortfree}.
For sort-free compositing more broadly, weighted blended order-independent
transparency~\cite{mcguire2013weighted} provides a useful conceptual analogue,
although it is not itself a Gaussian rendering method.
These approaches are largely orthogonal to TileGS when they change the
representation, compositing rule, or primitive set rather than the tile-local
traversal structure.
TileGS keeps the projected Gaussian inputs and splat evaluation equations
unchanged and instead reorganizes how the surviving Gaussians are traversed
during rasterization---a raster-pipeline redesign rather than a new scene
representation, training objective, or compression method.

Closest to TileGS are methods that revisit sorting and rendering efficiency in
3DGS.
StopThePop improves view consistency through hierarchical resorting and
culling~\cite{radl2024stopthepop};
sort-free rendering replaces alpha blending with weighted-sum rendering to avoid
depth sorting~\cite{hou2025sortfree};
stochastic splatting removes deterministic sorting through stochastic
rasterization and Monte Carlo estimation~\cite{kheradmand2025stochasticsplats};
and systems such as FlashGS and Speedy-Splat pursue broader speedups through
renderer-level optimization, scheduling, tighter Gaussian--tile localization, or
primitive sparsification~\cite{feng2025flashgs,hanson2024speedysplat}.
FlashGS in particular targets broader renderer efficiency through redundancy
elimination, pipelining, and memory-access scheduling.
TileGS instead isolates a narrower traversal-organization change: it keeps the
Gaussian set and per-Gaussian evaluation unchanged, but changes the tile-local
granularity at which assigned Gaussians are consumed.
The two directions therefore address different levels of the rasterization
pipeline and are plausibly complementary rather than competing.
TileGS differs by keeping the same Gaussian set and alpha-compositing target,
while changing only the tile-local traversal structure and selectively repairing
ambiguous local segments (Table~\ref{tab:method_positioning}).

\begin{table}[tbp]
  \centering
  \caption{Conceptual positioning of TileGS.}
  \label{tab:method_positioning}
  \setlength{\tabcolsep}{3pt}
  \renewcommand{\arraystretch}{1.05}
  \small
  \begin{tabular}{p{0.34\columnwidth}p{0.56\columnwidth}}
    \toprule
    Direction & Difference from TileGS \\
    \midrule
    Sort-free / stochastic &
    Change compositing or sampling; TileGS keeps alpha compositing. \\
    Pruning / sparsification &
    Reduce the Gaussian set; TileGS keeps the same workload. \\
    Broad renderer systems &
    Combine many optimizations; TileGS isolates tile-local bins plus repair. \\
    gsplat-style backends &
    Optimize the standard stream; TileGS changes traversal granularity. \\
    \bottomrule
  \end{tabular}
\end{table}

\subsection{Tile-based rendering and TBDR}

Tile-based rendering has long been used to improve locality and reduce external
memory traffic by organizing work around image-space tiles.
In Tile-Based Deferred Rendering (TBDR), primitives are first assigned to tiles,
and each tile is then processed largely within fast on-chip storage before the
final color results are written back to DRAM.
This execution model is especially important on mobile and SoC-class GPUs,
where off-chip bandwidth is comparatively scarce and energy-expensive.
A renderer may be tile-based in a broad sense while still exposing highly
irregular memory accesses if the execution order within each tile is poorly
matched to the underlying data layout --- a distinction central to this paper.
TileGS is inspired by clustered and bucketed rendering.
Clustered deferred and forward shading groups view samples into spatial/depth
clusters rather than only 2D tiles~\cite{olsson2012clustered}, while
order-independent transparency methods such as bucket depth peeling and layered
weighted blended OIT use depth buckets or layers to avoid globally sorting all
fragments~\cite{liu2009bucket,friederichs2021layered}.
TileGS adapts this idea to 3D Gaussian Splatting by forming depth-local ranges
inside each image tile and using selective repair to preserve front-to-back
Gaussian compositing, addressing the mismatch between globally assembled
traversal order and tile-local execution that the standard rasterizer leaves
unresolved.

\section{TileGS pipeline}
\label{sec:method}

TileGS replaces the globally sorted tile-depth stream traversal of the baseline
3DGS rasterizer with a tile-local depth-binned execution pipeline, without
changing the Gaussian representation, projected inputs, or compositing semantics.
The key design choice is to separate execution order from rendering semantics:
coarse bins reorganize traversal for performance, while selective repair restores
the ordering needed for baseline-equivalent compositing.
Rather than treating each tile as one long depth-sorted range, TileGS partitions
Gaussian entries into a small number of coarse depth bins, traverses them in
front-to-back order, and repairs only the segments where coarse local ordering is
insufficient.

\begin{figure*}[t]
  \centering
  \includegraphics[width=0.98\textwidth,height=0.33\textheight,keepaspectratio]{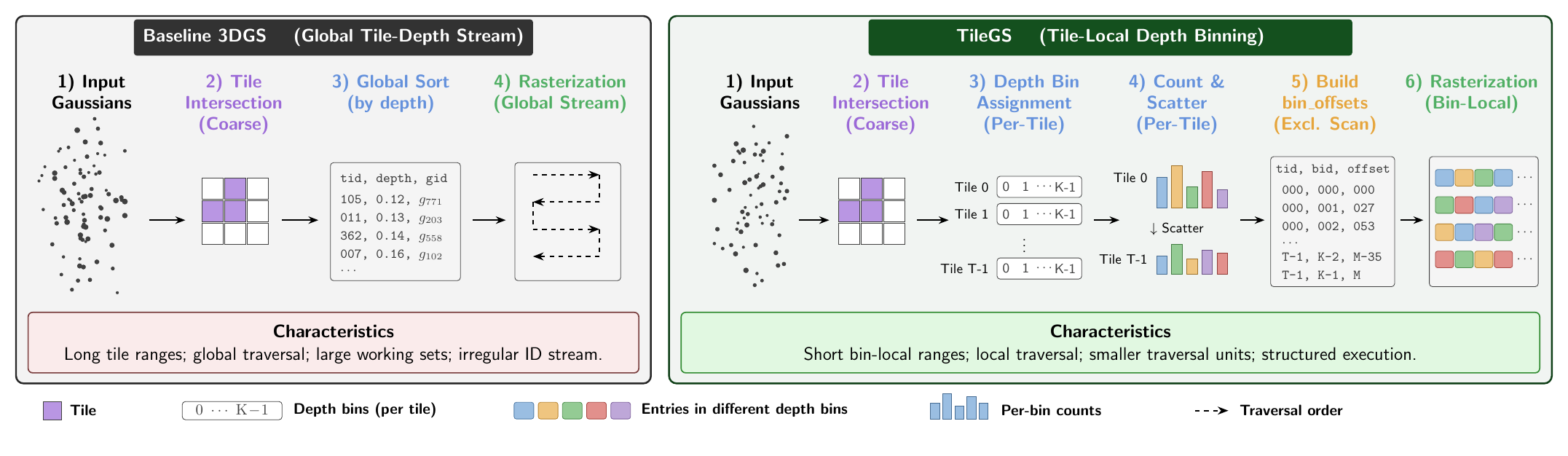}
  \caption{
    Pipeline comparison between baseline 3DGS rasterization and TileGS.
    The baseline constructs a globally sorted tile-depth stream, so each tile is
    rasterized as one long globally sorted tile-depth range.
    TileGS reorganizes the same Gaussian--tile entries into tile-local depth bins
    (steps 1--6), builds a bin-offset array via count-and-scatter and prefix sum,
    and rasterizes the resulting tile-bin-major stream in step 6.
    The default raster kernel consumes each tile's concatenated bin span linearly;
    bin boundaries determine the materialized stream layout and repair segments
    rather than separate raster-loop exits.
  }
  \label{fig:pipeline_overview}
\end{figure*}

Figure~\ref{fig:pipeline_overview} summarizes the pipeline.
Starting from projected Gaussians and their tile intersections, TileGS estimates
a robust depth range, assigns each Gaussian--tile entry to a coarse depth bin,
counts and scatters entries by $(\mathrm{tile}, \mathrm{bin})$ key, builds the
corresponding bin-offset structure, and rasterizes each tile by walking its bins
in order.
Repair handles the subset of cases in which coarse binning alone is not
sufficient to recover the exact baseline ordering.

\noindent\textbf{Terminology.}
We refer to the standard gsplat forward path as the \emph{globally sorted
tile-depth stream traversal}.
Our default method is \emph{TileGS No-GW} (\emph{No Geometry-Write}), which
means that rasterization consumes the tile-local binned Gaussian-ID stream
directly, without an additional pre-raster geometry materialization step.
We use \emph{TileGS Packed-GW} to denote a bandwidth-oriented variant that
inserts such a geometry-write stage and stores a more compact sidecar
representation for later raster traversal.
In short, No-GW prioritizes inner-loop simplicity and lower raster traversal
cost, whereas Packed-GW prioritizes reduced geometry traffic.

\subsection{Baseline 3DGS rasterization}
\label{sec:baseline}

The baseline 3DGS forward path projects each Gaussian to screen space,
determines the image tiles intersected by its footprint, and emits one
tile-intersection record per covered tile.
These records are then globally sorted by a composite key that first groups by
tile and then orders by depth.
After sorting, each tile is associated with one contiguous range in the global
intersection list, and the raster kernel traverses this range to evaluate
Gaussian contributions for the pixels of that tile.

This execution model is efficient in aggregate and serves as a strong baseline.
In particular, we compare against \emph{gsplat}, a widely used high-performance
implementation of Gaussian splatting, rather than the original vanilla 3DGS
reference code.

Although both gsplat and TileGS are tile-based, they differ in how tile work is
organized. gsplat assembles Gaussian--tile intersections into a globally sorted
tile-depth stream, so each tile rasterizes one contiguous front-to-back range.
TileGS keeps the same projected Gaussians, tile assignment, and compositing
equations, but decomposes each tile's range into short depth-local bins and
applies selective exact repair to high-risk segments. Thus, the novelty is not
tiling itself, but replacing globally sorted tile-depth stream traversal with
tile-local depth-binned traversal plus targeted repair.

Beyond this traversal change, TileGS adds three auxiliary stages absent from the
gsplat forward path: depth-range estimation (\S\ref{sec:depth_range}),
tile-local bin construction (\S\ref{sec:bin_construction}), and selective repair
(\S\ref{sec:repair}). All reported end-to-end frame times include these stages.

Our instrumentation shows severe scatter on representative scenes: the mean
adjacent-lane index difference $|\Delta g|$ is on the order of 1.4M, the mean
warp $g$-span reaches about 4.9M, and only about 0.4\% of warps exhibit
monotone or near-neighbor Gaussian-index sequences.
Importantly, TileGS does not significantly change these Gaussian-index stream
(g-stream) scatter statistics:
under the No-GW binned path, mean $|\Delta g|$ decreases by only $1$--$3\%$
and the monotonic warp fraction remains below $0.4\%$.
TileGS improves raster performance by changing traversal granularity rather
than memory-access locality or coalescing.
Shorter per-bin working sets and a more structured raster execution pattern
allow the kernel to complete in fewer elapsed cycles without materially changing
the underlying scatter pattern, as we quantify in \S\ref{sec:analysis}.

\subsection{Depth range estimation}
\label{sec:depth_range}

TileGS assigns each Gaussian--tile entry to a coarse depth bin before bin
construction.
To make this mapping stable across scenes and viewpoints, we first estimate a
robust effective depth range for the current frame and then normalize projected
Gaussian depths within that range.

A direct min--max normalization is undesirable because projected depth
distributions can contain strong outliers.
Rare extreme depths would stretch the range and waste most bins on depths that
contribute little to the visible tile workload.
Instead, TileGS samples visible projected depths and selects an effective
interval $[z_{\min}, z_{\max}]$ from clipped low and high percentiles, followed
by a small padding margin.
In the reported implementation, we sample up to 8192 depth candidates and select
the 1st/99th percentiles using exact order statistics (\texttt{kthvalue}).
A 128-bin log-spaced histogram estimator exists as an unused alternative, but
all reported results use the exact-order-statistics path.

Within this clipped range, the current implementation uses \emph{log-based}
depth binning rather than a uniform linear mapping.
The reason is that near-depth regions generally require finer separation:
Gaussians closer to the camera tend to occupy larger footprints, overlap more
strongly in image space, and are more sensitive to coarse ordering errors.
Log binning therefore allocates more resolution near the camera while keeping
the total number of bins moderate.

This stage is only a coarse local ordering device.
Its purpose is to create shorter and more structured tile-local traversal
ranges, not to recover exact compositing order by itself.
Any remaining exactness requirement is handled later by the repair stage.
Accordingly, the depth-to-bin mapping should be understood as a performance
device rather than a semantic requirement.

\subsection{Tile-local bin construction}
\label{sec:bin_construction}

After assigning each Gaussian--tile entry to a coarse depth bin, TileGS
replaces the single globally sorted tile-depth stream with a tile-local bin
construction pass.
Each entry is associated with a composite key
$(\mathrm{tile\_id}, \mathrm{bin\_id})$, where $\mathrm{tile\_id}$ identifies
the covered image tile and $\mathrm{bin\_id}$ is obtained from the normalized
depth mapping described above.
The purpose of this stage is to reorganize the tile workload into smaller
front-to-back ranges that can later be traversed independently by the raster
kernel.
This reorganization does not reduce the number of Gaussian--tile entries
assigned to a tile; it only changes their grouping and front-to-back order
(the static/dynamic distinction is formalized in \S\ref{sec:analysis}).

\begin{figure*}[t]
  \centering
  \includegraphics[width=0.98\textwidth,height=0.33\textheight,keepaspectratio]{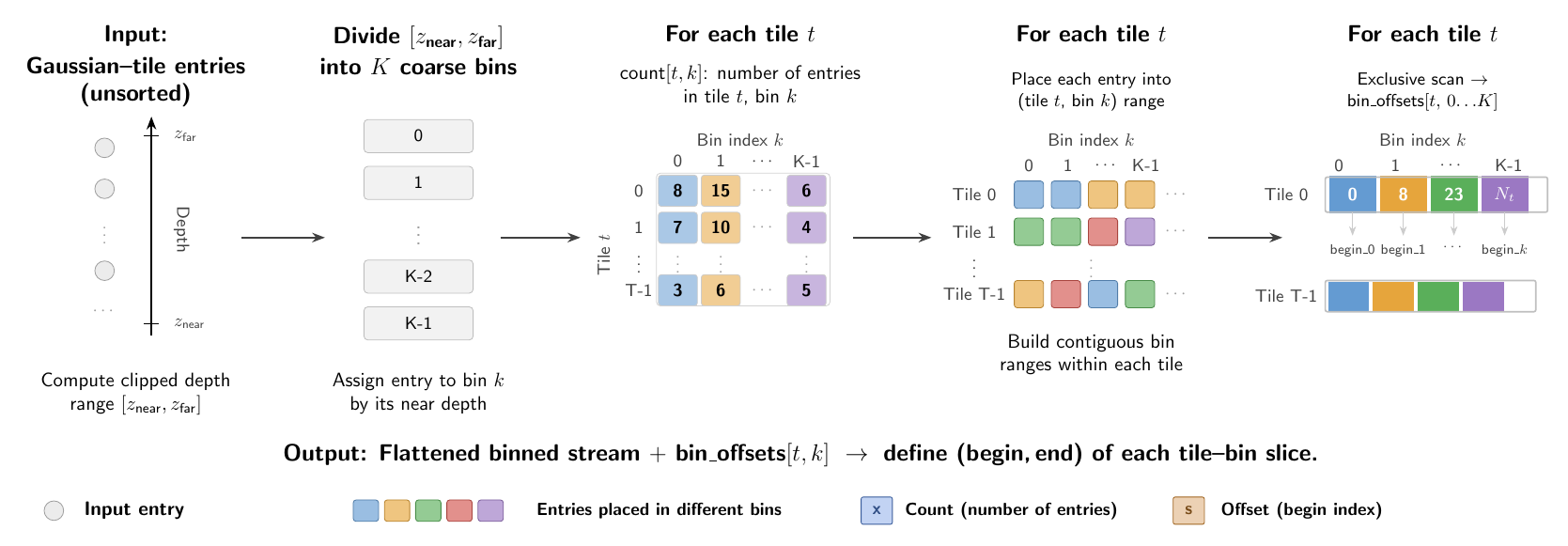}
  \caption{
    \textbf{Tile-local bin construction.}
    TileGS assigns Gaussian--tile entries to coarse depth bins,
    counts per-bin populations, scatters entries into contiguous
    bin ranges, and builds the bin-offset array via exclusive scan.
  }
  \label{fig:bin_construction}
\end{figure*}

The construction proceeds in four stages, illustrated in
Figure~\ref{fig:bin_construction}.
First, TileGS counts the number of entries assigned to each
$(\mathrm{tile}, \mathrm{bin})$ pair, producing a tile-local bin histogram.
Second, an exclusive prefix sum over this histogram yields the bin-offset array,
which defines the contiguous storage interval of every bin in the flattened
tile-bin stream.
Third, the Gaussian--tile entries are scattered into these intervals.
Finally, the resulting flat list of Gaussian IDs is materialized in the order
expected by the binned rasterizer.

Critically, this flattened stream is laid out in \emph{tile-bin-major} order:
the bin-offset array groups bins by tile first and orders bins front-to-back
within each tile, so every bin belonging to a given tile occupies one
contiguous sub-range of the stream, and those sub-ranges themselves sit back
to back within one larger per-tile span.
As a result, each tile's raster kernel still walks exactly one contiguous
interval of the flattened stream---matching the baseline's per-tile contiguous
access pattern---but that interval is now pre-organized into contiguous,
depth-ordered bin sub-ranges rather than being only the baseline's single
per-tile depth-sorted list.
This tile-bin-major layout is what the binned raster kernel walks directly;
bin boundaries are materialized during construction and repair, while the
default raster traversal consumes the final per-tile span linearly.

\noindent\textbf{Count/scatter implementation variants.}
We evaluated five aggregation strategies for the tile-local count/scatter pass
(direct atomics, warp-private histograms, two-stage aggregation, block-local
hash aggregation, and warp-key grouped scatter); all produce identical
bin-offset arrays and Gaussian-ID streams and therefore do not affect raster
semantics, repair behavior, or image quality.
In the final configuration, we use the direct-atomic construction path with
warp-key grouping disabled.
In the completed full-9 RTX~4090 variant sweep, warp-private histograms were
slower than the default on every scene and 16.0\% slower on average; a later
paired rerun also favored warp-key grouping disabled by a small median margin
with a tighter run range.
Detailed timings are provided in the supplement.
We treat the selected count/scatter strategy as an implementation-level detail
rather than an algorithmic contribution of TileGS.

\subsection{Binned rasterization: No-GW and Packed-GW}

We study two execution styles built on the same tile-local binning framework.
No-GW consumes the binned Gaussian-ID stream directly and fetches geometry
attributes from the original arrays. Packed-GW inserts an additional pre-raster
geometry-write stage that materializes a compact sidecar representation for
later raster traversal.

Packed-GW is best read as an ablation that isolates the cost of geometry-write
materialization: it changes bandwidth characteristics without changing which
Gaussians are consumed or the tile-bin-major traversal and repaired ordering, so
it does not represent a distinct rasterization algorithm from No-GW.

Empirically, No-GW gives the strongest end-to-end result because it avoids the
extra pre-raster materialization stage. Packed-GW is therefore treated as a
bandwidth-oriented diagnostic rather than the default method: a profiling sweep
shows reduced raster DRAM traffic in 7/9 scenes, but a follow-up timing-only
rerun is slower on all nine scenes. Details are provided in the supplement.

\subsection{Repair for correctness}
\label{sec:repair}

Coarse tile-local depth binning is sufficient to expose a more structured raster
workload, but it is not, by itself, a guarantee of exact front-to-back
compositing order.
Two Gaussians that fall into the same coarse bin, or that straddle a bin
boundary under the normalized depth mapping, may still require a finer ordering
relationship than the bin traversal alone provides.
TileGS therefore augments binned rasterization with a repair stage that restores
the baseline compositing behavior only where the coarse local ordering is
insufficient.

\begin{figure*}[t]
  \centering
  \includegraphics[width=0.98\textwidth,height=0.33\textheight,keepaspectratio]{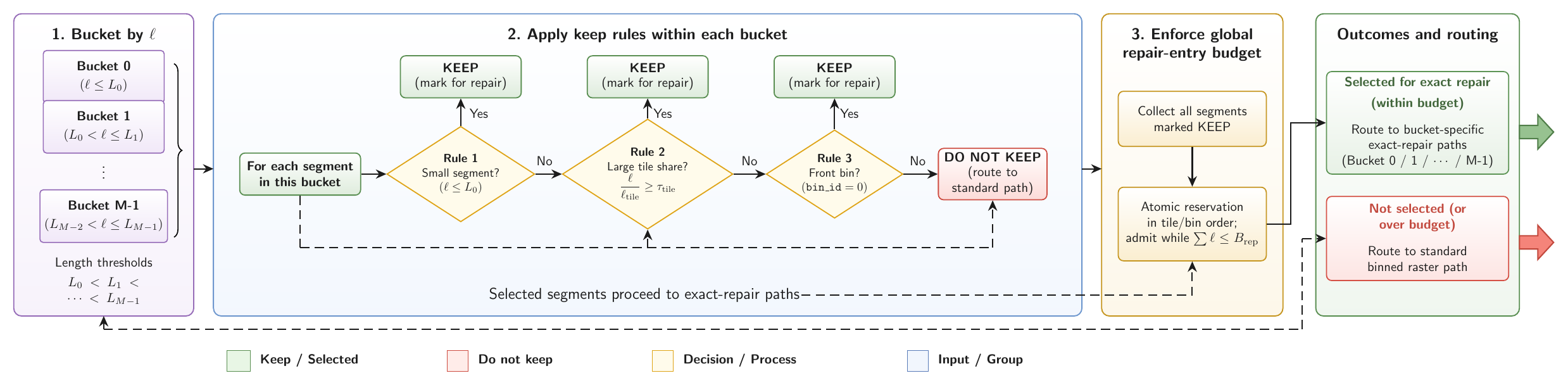}
  \caption{
    \textbf{Selection of exact-repair segments.}
    TileGS treats each coarse tile/bin slice as a repair candidate, buckets
    segments by length, applies keep rules based on segment size, tile share,
    and front-bin priority, and then enforces a global repair-entry budget via
    atomic reservation in tile/bin index order.
    Selected segments are routed to bucket-specific exact-repair paths, while
    non-selected segments continue through the standard binned raster path.
  }
  \label{fig:repair_selection}
\end{figure*}

The repair logic is segment-based.
Each logical segment corresponds to one coarse tile/bin slice
$(\mathrm{tile\_id}, \mathrm{bin\_id})$ in the flattened binned stream, with
segment length
\[
\ell = \mathrm{seg\_end} - \mathrm{seg\_begin}.
\]
Segments with $\ell \le 1$ are treated as not requiring repair in the current
implementation.
Longer segments are bucketed by length into local exact-repair buckets
($2\!-\!128$, $129\!-\!256$, $257\!-\!512$) and a global-tail bucket
($>512$), with the default local exact-repair limit set to 512 entries
(Figure~\ref{fig:repair_selection}).
In the implementation the $2\!-\!128$ range is further sub-divided into
$2\!-\!32$, $33\!-\!64$, and $65\!-\!128$ sub-buckets, each dispatched to a
fixed-size bitonic sort kernel; the paper groups these as one bucket for
brevity.

In the current implementation, exact-repair selection uses the following
default policy.
Let $\ell$ be the segment length and $\ell_{\mathrm{tile}}$ be the total number
of entries in the parent tile.
Table~\ref{tab:repair_rules} summarizes the five selection rules and their
rationale.

\begin{table}[tbp]
  \centering
  \caption{%
    Exact-repair selection rules.
    A segment is marked for repair if \emph{any} rule fires.
    $\ell$: segment length; $\ell_{\mathrm{tile}}$: total entries in the parent
    tile; bin\,0/1 refers to the two nearest-camera depth bins.
    All thresholds are fixed global hyperparameters.%
  }
  \label{tab:repair_rules}
  \setlength{\tabcolsep}{4pt}
  \renewcommand{\arraystretch}{1.08}
  \small
  \begin{tabular}{clp{0.44\columnwidth}}
    \toprule
    Rule & Condition & Rationale \\
    \midrule
    (i)   & $\ell \ge 320$
          & Long segments likely affect visible pixels regardless of opacity. \\
    (ii)  & $\ell > 512$ (global-tail bucket)
          & Same motivation; forces repair for all tail-bucket segments. \\
    (iii) & $\ell / \ell_{\mathrm{tile}} \ge 45\%$
          & Segment dominates the tile; ordering errors concentrate at one depth slice. \\
    (iv)  & bucket $129$--$256$ \& $\ell / \ell_{\mathrm{tile}} \ge 10\%$
          & Medium-length segment with significant tile share. \\
    (v)   & bin\,id $\in \{0,1\}$ \& $\ell \ge 16$
          & Near-camera Gaussians are large, high-opacity, and most order-sensitive. \\
    \bottomrule
  \end{tabular}
\end{table}

A segment is selected if any rule in Table~\ref{tab:repair_rules} fires.
Rules (i)--(ii) target long segments, rules (iii)--(iv) target segments that
occupy a large fraction of their parent tile, and rule (v) prioritizes
near-camera bins where ordering errors are most visible.
Each rule follows directly from the compositing-error argument above: segments
are repaired when they are most likely to contain Gaussians with non-negligible
opacity at non-negligible transmittance.
All thresholds are fixed global hyperparameters, not tuned per scene or platform.

\noindent\textbf{Exact repair.}
Exact repair operates only on the selected ambiguous local segments.
Within such a segment, TileGS reconstructs the order needed to match the
baseline raster semantics and replays that segment in the corrected order.
This correction is strictly local rather than global: only the selected segment
is repaired, while the rest of the tile continues to use the standard binned
path.

\noindent\textbf{Adaptive repair.}
The repair policy is adaptive in two senses.
First, different segment-length buckets map to different local exact-repair
paths, with small and medium segments handled entirely in local memory and
larger segments using progressively broader exact paths.
Second, the selected repair set is globally budgeted, as described above, so
large or numerous ambiguous segments cannot dominate the frame.

\noindent\textbf{Correctness statement.}
After repair, TileGS matches the baseline output up to numerical noise on all
reported experiments ($|\Delta\mathrm{PSNR}| < 0.001~\mathrm{dB}$,
$|\Delta\mathrm{SSIM}| < 0.001$, and
$|\Delta\mathrm{LPIPS}| < 0.001$).
The repaired path should therefore be interpreted as empirically
equivalent to the baseline within numerical precision under the reported
benchmarks, rather than as a deliberately quality-degrading approximation.

\noindent\textbf{Why partial repair suffices.}
A natural question is why a budget-constrained, threshold-based repair policy
achieves near-exact output when the majority of segments are left on the
standard binned path.

Partial repair is effective because most local ordering errors have limited
image impact. In alpha compositing, the color change from swapping two adjacent
Gaussians is suppressed when either opacity is small or accumulated
transmittance is already low. Shallow depth bins reduce the set of possible
misorderings, but do not by themselves guarantee exact output because the
effect also depends on opacity, color contrast, and transmittance.

TileGS therefore treats coarse binning as an execution-order optimization, not
as a correctness argument. Exact repair is applied only to high-risk tile/bin
segments selected by the policy above, while short or low-impact segments remain
on the standard binned path.

\section{Experimental setup}
\label{sec:setup}
We evaluate TileGS across representative 3DGS benchmark scenes and hardware
platforms under controlled measurement settings. All primary comparisons use the
same projected Gaussian inputs and compare the standard gsplat globally sorted
tile-depth stream traversal against TileGS No-GW.

\subsection{Hardware platforms and software stack}

We evaluate TileGS on two NVIDIA Ada GPUs representing distinct operating
regimes: a desktop-class RTX~4090 and a laptop-class RTX~1000~Ada.
All measurements use controlled clock presets to reduce run-to-run variability
from dynamic boost, thermal throttling, and power-management behavior.
These presets fix the compute-to-bandwidth operating point for each platform,
so cross-platform comparisons should be interpreted as controlled measurements
under the stated presets rather than as claims about all possible boost states.
Table~\ref{tab:platforms} summarizes the hardware and software configuration.
Both platforms use the same gsplat commit \texttt{a8d88d3}.

\begin{table}[tbp]
  \centering
  \caption{Evaluation platforms under controlled clock presets.}
  \label{tab:platforms}
  \setlength{\tabcolsep}{4pt}
  \begin{tabular}{lcc}
    \toprule
    & RTX~4090 & RTX~1000~Ada \\
    \midrule
    Platform class & Desktop Ada & Laptop Ada \\
    Clock preset & 2100/10501\,MHz & 1200/6001\,MHz \\
    Power limit & 350\,W & controlled laptop preset \\
    PyTorch & 2.6.0+cu124 & 2.8.0+cu128 \\
    CUDA / driver & 12.2 / 535.288.01 & 13.1 / 591.86 \\
    Nsight Compute & 2026.1.0.0 & 2023.2.1.0 \\
    gsplat commit & \multicolumn{2}{c}{\texttt{a8d88d3}} \\
    \bottomrule
  \end{tabular}
\end{table}

\subsection{Benchmark scenes}
We evaluate on a 9-scene benchmark drawn from three standard 3DGS evaluation
datasets: Mip-NeRF~360~\cite{barron2022mipnerf360},
Tanks and Temples~\cite{knapitsch2017tanks}, and
Deep Blending~\cite{hedman2018deepblending}, summarized in
Table~\ref{tab:benchmark_scenes}.
The scenes span raster-heavy outdoor cases such as \textit{bicycle},
\textit{garden}, and \textit{stump}, where long tile ranges and high overlap
stress the raster kernel, as well as lighter or more compact scenes such as
\textit{kitchen}, \textit{bonsai}, \textit{drjohnson}, and \textit{playroom},
where auxiliary binning and repair overheads are more visible.
Together, this benchmark evaluates both the raster-heavy regime where TileGS is
expected to benefit most and the lighter-scene regime where a tile-local
pipeline must avoid regressions.

\begin{table}[tbp]
  \centering
  \caption{
    Benchmark scenes used in the evaluation.
    The 9-scene suite follows the common 3DGS/gsplat evaluation split and covers
    outdoor, indoor, object-centric, and complex blended scenes.
  }
  \label{tab:benchmark_scenes}
  \setlength{\tabcolsep}{6pt}
  \begin{tabular}{ll}
    \toprule
    Dataset & Scenes \\
    \midrule
    Mip-NeRF 360 &
    \textit{bicycle}, \textit{garden}, \textit{stump},
    \textit{kitchen}, \textit{bonsai} \\
    Tanks and Temples &
    \textit{train}, \textit{truck} \\
    Deep Blending &
    \textit{drjohnson}, \textit{playroom} \\
    \bottomrule
  \end{tabular}
\end{table}

\subsection{Baselines and variants}

Our primary baseline is \textbf{gsplat}~\cite{ye2024gsplat}, a modern
high-performance 3DGS implementation that serves as a strong and fair
comparator for a rasterization pipeline redesign.

The primary comparison is between:
\begin{itemize}
  \item \textbf{Baseline (gsplat):} the standard optimized 3DGS rasterization
        pipeline with a globally sorted tile-depth stream; and
  \item \textbf{TileGS No-GW:} our default tile-local depth-binning pipeline
        with $K=64$.
\end{itemize}

Unless otherwise stated, all main results use TileGS No-GW, which is the
strongest end-to-end variant in the current implementation; Packed-GW is
included only as a bandwidth-oriented ablation.

\subsection{Metrics and profiling methodology}
\label{sec:metrics}

Our primary system metric is end-to-end frame time in milliseconds.
This covers the full GPU forward pass — spherical-harmonics evaluation,
projection, tile intersection, sorting, rasterization, and all TileGS auxiliary
stages (depth-range estimation, bin construction, repair) — timed via
CUDA events on the device side.
It excludes Python-side host overhead, CPU-to-GPU data transfer of the Gaussian
parameters (which are assumed resident), and the backward pass.
To isolate the effect of the raster pipeline itself, we also measure the runtime
of the main raster kernel.
This distinguishes the speedup due to binned raster traversal from the total
frame-time impact of auxiliary stages such as depth-range estimation, bin
construction, sidecar generation, and repair.

Image quality is measured with PSNR, SSIM, and LPIPS against the gsplat output.
TileGS matches the baseline up to numerical noise
($|\Delta\mathrm{PSNR}| < 0.001~\mathrm{dB}$,
$|\Delta\mathrm{SSIM}| < 0.001$, and
$|\Delta\mathrm{LPIPS}| < 0.001$) on all reported scenes.

To understand the source of TileGS's speedup, we analyze DRAM traffic, cache
behavior, and effective throughput using Nsight Compute.
For source-attributed traffic analysis, we use theoretical L2 sector counts as a
proxy for memory pressure, since raw DRAM bytes are not directly attributable to
individual source lines in the profiler.
This allows us to separate total traffic from excess or uncoalesced geometry
loads.

\subsection{Implementation settings}

Unless otherwise stated, the default TileGS configuration uses:
\begin{itemize}
  \item \textbf{No-GW} as the primary raster path;
  \item \textbf{$K=64$} depth bins;
  \item \textbf{direct-atomic count/scatter} construction, with warp-key grouping disabled;
  \item conservative tile alpha culling in the raster kernel; and
  \item the repair path for exact output recovery.
\end{itemize}

The choice of $K=64$ is based on the ablation study (\S\ref{sec:results_kbins}),
where it provides the best overall fixed setting, with $K=32$ remaining a strong
near-optimal alternative.

\section{Results}
\label{sec:results}

We first report the main end-to-end comparison against the gsplat baseline,
then isolate the rasterization-kernel speedup, summarize the depth-bin ablation,
and finally confirm that TileGS preserves image quality.
Unless otherwise stated, all results in this section use the default No-GW
configuration with $K=64$.

\subsection{End-to-end frame time}
\label{sec:results_frame}

\begin{table}[tbp]
  \centering
  \caption{
    End-to-end frame time (ms) and speedup for Baseline vs.\ TileGS No-GW
    on RTX~4090 and RTX~1000~Ada under locked presets.
  }
  \label{tab:main_results}
  \setlength{\tabcolsep}{5pt}
  \begin{tabular}{lcccccc}
    \toprule
    & \multicolumn{3}{c}{\textbf{RTX~4090}} & \multicolumn{3}{c}{\textbf{RTX~1000~Ada}} \\
    \cmidrule(lr){2-4} \cmidrule(lr){5-7}
    Scene & Base & TileGS & $\times$ & Base & TileGS & $\times$ \\
    \midrule
    \textit{bicycle}   & 4.395 & 3.979 & 1.105 & 38.92 & 34.78 & 1.119 \\
    \textit{garden}    & 3.998 & 3.602 & 1.110 & 36.85 & 32.33 & 1.140 \\
    \textit{stump}     & 2.957 & 2.737 & 1.080 & 27.30 & 24.33 & 1.122 \\
    \textit{kitchen}   & 2.638 & 2.476 & 1.065 & 25.59 & 22.03 & 1.162 \\
    \textit{bonsai}    & 1.649 & 1.525 & 1.081 & 13.35 & 12.32 & 1.084 \\
    \textit{train}     & 2.257 & 2.145 & 1.052 & 17.96 & 17.32 & 1.037 \\
    \textit{truck}     & 2.523 & 2.427 & 1.039 & 23.13 & 21.59 & 1.072 \\
    \textit{drjohnson} & 2.348 & 2.224 & 1.056 & 21.16 & 20.07 & 1.054 \\
    \textit{playroom}  & 1.980 & 1.919 & 1.032 & 18.21 & 17.25 & 1.056 \\
    \midrule
    \textbf{Mean} & & & \textbf{1.069} & & & \textbf{1.094} \\
    \bottomrule
  \end{tabular}
\end{table}

Table~\ref{tab:main_results} shows that TileGS No-GW improves all nine scenes,
with mean end-to-end speedups of 1.069$\times$ on RTX~4090 and 1.094$\times$
on RTX~1000~Ada.
The larger laptop-GPU gain is consistent with a more bandwidth-constrained
operating point.

The largest gains occur on raster-heavy scenes such as \textit{bicycle},
\textit{garden}, \textit{stump}, and \textit{kitchen}, where long raster ranges
constitute a larger fraction of total cost and better amortize depth-range
estimation, bin construction, and repair.
End-to-end speedups exceed 10\% on both platforms for several such scenes, while
the lower scene-averaged mean reflects lighter cases where auxiliary overhead is
proportionally more visible.
TileGS improves all nine scenes on both GPUs under locked clock settings.

\subsection{Rasterization kernel speedup}
\label{sec:results_kernel}

\begin{table}[tbp]
  \centering
  \caption{
    Rasterization-kernel speedup and frame/kernel gap.
    Gap = (kernel speedup $-$ frame speedup) in percentage points.
    Missing RTX~1000~Ada kernel entries are profiler-capture failures, not
    application failures; the corresponding end-to-end timings are reported in
    Table~\ref{tab:main_results}.
    \textsuperscript{\dag} Ada mean is computed over the five scenes for which
    full-resolution kernel captures completed.
  }
  \label{tab:kernel_speedup}
  \setlength{\tabcolsep}{5pt}
  \begin{tabular}{lcccccc}
    \toprule
    & \multicolumn{3}{c}{\textbf{RTX~4090}} & \multicolumn{3}{c}{\textbf{RTX~1000~Ada}} \\
    \cmidrule(lr){2-4} \cmidrule(lr){5-7}
    Scene & Kernel & Frame & Gap & Kernel & Frame & Gap \\
    \midrule
    \textit{bicycle}   & 1.422 & 1.105 & 31.7 & ---   & 1.119 & --- \\
    \textit{garden}    & 1.456 & 1.110 & 34.6 & ---   & 1.140 & --- \\
    \textit{stump}     & 1.426 & 1.080 & 34.6 & ---   & 1.122 & --- \\
    \textit{kitchen}   & 1.430 & 1.065 & 36.5 & 1.439 & 1.162 & 27.7 \\
    \textit{bonsai}    & 1.450 & 1.081 & 36.9 & 1.441 & 1.084 & 35.7 \\
    \textit{train}     & 1.452 & 1.052 & 40.0 & 1.448 & 1.037 & 41.1 \\
    \textit{truck}     & 1.511 & 1.039 & 47.2 & 1.457 & 1.072 & 38.5 \\
    \textit{drjohnson} & 1.436 & 1.056 & 38.0 & ---   & 1.054 & --- \\
    \textit{playroom}  & 1.368 & 1.032 & 33.6 & 1.418 & 1.056 & 36.2 \\
    \midrule
    \textbf{Mean} & \textbf{1.439} & & & \textbf{1.441}\textsuperscript{\dag} & & \\
    \bottomrule
  \end{tabular}
\end{table}

Table~\ref{tab:kernel_speedup} isolates the main rasterization-kernel speedup.
We note upfront that detailed kernel captures on RTX~1000~Ada were unavailable
for \textit{bicycle}, \textit{garden}, \textit{stump}, and \textit{drjohnson}
due to driver/profiler resource acquisition errors (the application runs
completed normally, and all nine end-to-end timings are intact in
Table~\ref{tab:main_results}).
Because those four scenes include some of the heaviest outdoor workloads where
TileGS achieves its largest end-to-end gains, we treat the resulting Ada kernel
mean as a five-scene subset rather than a complete full-suite average, and we do
not use it to claim architecture-invariant kernel speedup.
On RTX~4090, where all nine captures succeeded, TileGS achieves a mean
raster-kernel speedup of 1.439$\times$; the Ada five-scene subset yields
1.441$\times$, consistent with the same kernel-level trend on the primary
platform.

Averaged over the available kernel captures, TileGS reaches a
1.439$\times$ raster-kernel speedup on RTX~4090 and 1.441$\times$
on RTX~1000~Ada, while the corresponding end-to-end frame speedups are
1.069$\times$ and 1.094$\times$. This gap is not a contradiction:
TileGS substantially accelerates the raster kernel, but the current
implementation spends part of that gain on bin construction and repair.
Per-stage profiling on \textit{bicycle} and \textit{garden}
(Table~\ref{tab:stage_breakdown} in \S\ref{sec:analysis_stages}) shows that
the global sort replacement saves 0.44--0.64\,ms, while bin construction
(0.40--0.44\,ms) and repair (0.33--0.41\,ms) are the dominant added costs.

\begin{table}[tbp]
  \centering
  \caption{
    Depth-bin count ablation on RTX~4090 over the full 9-scene benchmark.
    Quality deltas are measured relative to the $K=64$ TileGS configuration,
    not relative to the gsplat baseline.
    These intra-TileGS ablation deltas are smaller than the global
    TileGS-vs.-gsplat numerical-noise threshold used elsewhere.
  }
  \label{tab:kbins}
  \setlength{\tabcolsep}{4pt}
  \begin{tabular}{lccccc}
    \toprule
    K & Avg total & Avg frame & $|\Delta\mathrm{PSNR}|$ & $\Delta\mathrm{SSIM}$ & $\Delta\mathrm{LPIPS}$ \\
        & (ms) & (ms) & (dB) &  &  \\
    \midrule
    64  & \textbf{2.8787} & \textbf{3.0144} & 0 & 0 & 0 \\
    32  & 2.8973 & 3.0338 & $<10^{-4}$ & $<10^{-5}$ & $<10^{-5}$ \\
    128 & 2.9235 & 3.0641 & $<10^{-4}$ & $<10^{-5}$ & $<10^{-5}$ \\
    16  & 2.9357 & 3.0724 & $<10^{-4}$ & $<10^{-5}$ & $<10^{-5}$ \\
    8   & 2.9904 & 3.1255 & $<10^{-4}$ & $<10^{-5}$ & $<10^{-5}$ \\
    4   & 3.0985 & 3.2342 & $<10^{-4}$ & $<10^{-5}$ & $<10^{-5}$ \\
    \bottomrule
  \end{tabular}
\end{table}

\subsection{Depth bin count ablation}
\label{sec:results_kbins}

We ablate the number of depth bins $K$ in Table~\ref{tab:kbins}.
Very small values under-partition the tile stream, while very large values add
bookkeeping overhead. 
Across the full suite, $K=64$ gives the best fixed setting and $K=32$
remains a strong near-optimal alternative. 
Quality is invariant across the sweep: the maximum intra-TileGS drift remains below
$1.6\times10^{-5}$\,dB PSNR, $2.8\times10^{-6}$ SSIM, and
$1.7\times10^{-6}$ LPIPS.

\subsection{Image quality}

TileGS must not obtain speedups by degrading output.
Across the full 9-scene evaluation, TileGS No-GW matches gsplat within numerical
noise: $|\Delta\mathrm{PSNR}| < 0.001$~dB, $|\Delta\mathrm{SSIM}| < 0.001$, and
$|\Delta\mathrm{LPIPS}| < 0.001$ on all reported scenes.
Table~\ref{tab:quality} reports the absolute gsplat quality on RTX~4090, and
Figure~\ref{fig:quality_visual} shows a representative raster-heavy case.

\begin{table}[tbp]
  \centering
  \caption{
    Absolute image quality on RTX~4090. Values report the gsplat baseline;
    TileGS matches these values within $|\Delta\mathrm{PSNR}| < 0.001$~dB,
    $|\Delta\mathrm{SSIM}| < 0.001$, and $|\Delta\mathrm{LPIPS}| < 0.001$ on
    every scene.
  }
  \label{tab:quality}
  \setlength{\tabcolsep}{4pt}
  \begin{tabular}{@{}lrrr@{}}
    \toprule
    Scene & PSNR (dB) & SSIM & LPIPS \\
    \midrule
    \textit{bicycle}   & 23.65 & 0.687 & 0.213 \\
    \textit{garden}    & 25.22 & 0.769 & 0.129 \\
    \textit{stump}     & 26.26 & 0.750 & 0.173 \\
    \textit{kitchen}   & 25.44 & 0.833 & 0.105 \\
    \textit{bonsai}    & 27.56 & 0.890 & 0.125 \\
    \textit{train}     & 21.72 & 0.818 & 0.189 \\
    \textit{truck}     & 24.29 & 0.853 & 0.166 \\
    \textit{drjohnson} & 29.28 & 0.918 & 0.157 \\
    \textit{playroom}  & 29.82 & 0.920 & 0.153 \\
    \bottomrule
  \end{tabular}
\end{table}

\begin{figure*}[!t]
  \centering
  \setlength{\tabcolsep}{3pt}
  \begin{tabular}{ccc}
    \includegraphics[width=0.32\textwidth]{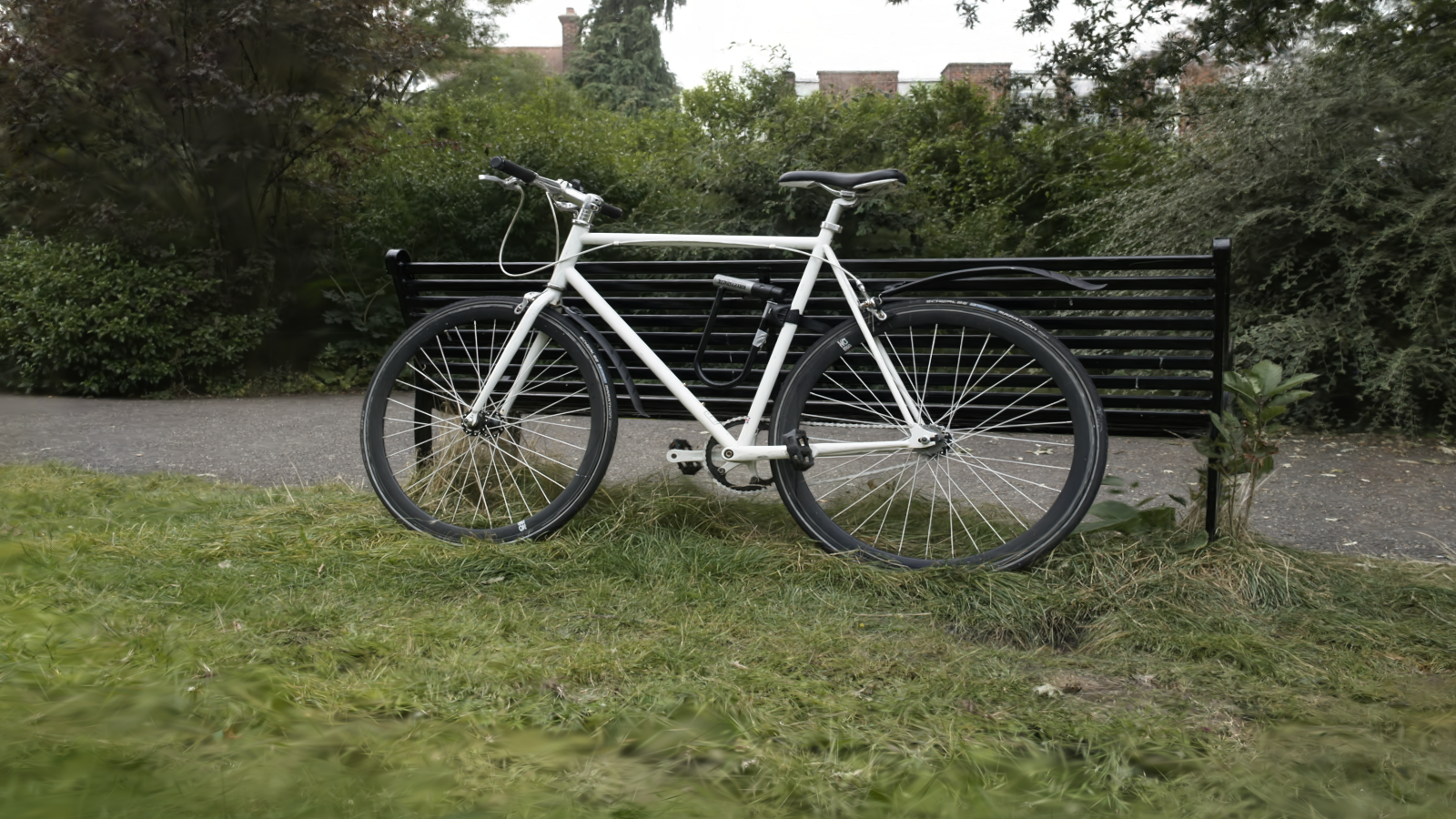} &
    \includegraphics[width=0.32\textwidth]{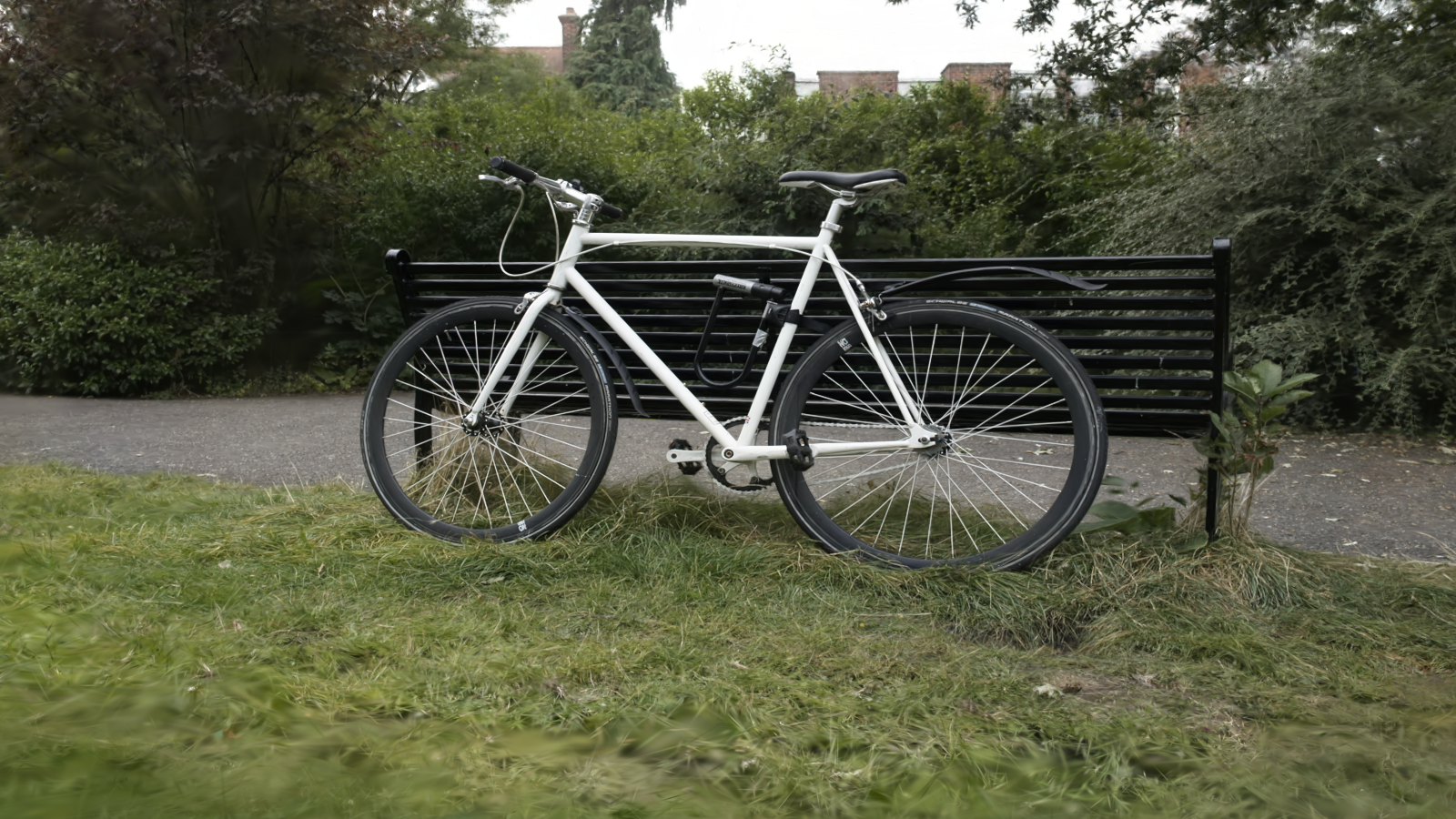} &
    \includegraphics[width=0.32\textwidth]{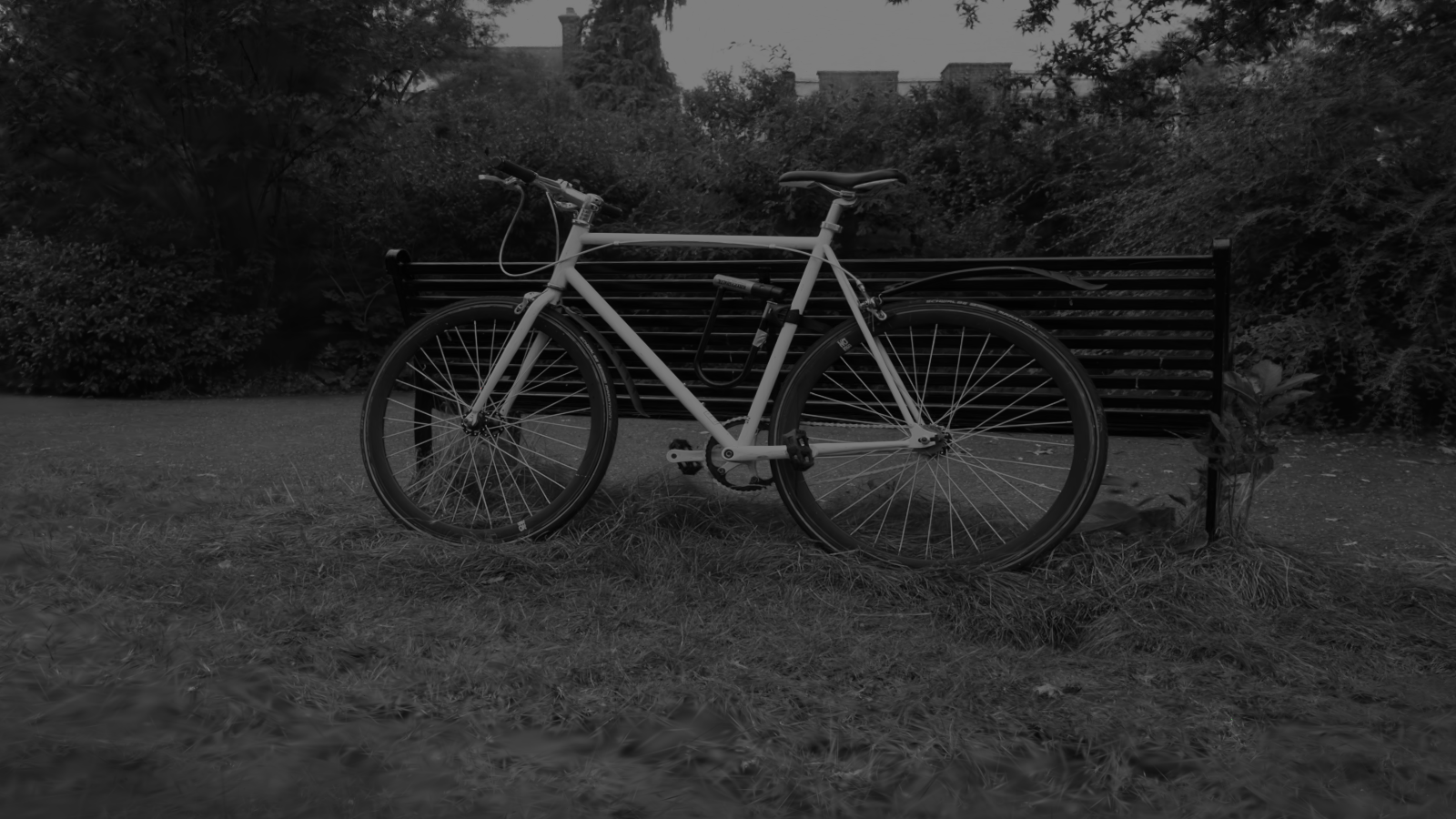} \\[2pt]
    \textbf{Baseline} & \textbf{TileGS} & \textbf{$100\times$ diff overlay}
  \end{tabular}
  \caption{
    Representative visual comparison on the raster-heavy \textit{bicycle} case.
    TileGS is visually indistinguishable from the gsplat baseline
    ($|\Delta\mathrm{PSNR}| < 0.001~\mathrm{dB}$ across all nine scenes);
    the rightmost column shows amplified absolute RGB differences overlaid
    on the TileGS render ($100\times$) for visibility.
  }
  \label{fig:quality_visual}
\end{figure*}

\subsection{Repair sensitivity}
\label{sec:results_repair}

To isolate the role of repair, Table~\ref{tab:repair_ablation} compares no
repair, the default selective-repair policy, and full repair on a representative
3-scene RTX~4090 subset spanning a raster-heavy scene (\textit{bicycle}), an
outdoor Tanks-and-Temples scene (\textit{truck}), and a lighter indoor scene
(\textit{playroom}).
Removing repair reduces frame time but breaks baseline equivalence, with a
maximum PSNR error of 2.03~dB and a maximum SSIM error of 0.0579.
In contrast, both the default policy and full repair match the baseline within
numerical noise.
Full repair does not improve quality meaningfully over the default policy on
this subset and is slower, indicating that the selective policy is sufficient
for the evaluated cases.

\begin{table}[tbp]
  \centering
  \setlength{\tabcolsep}{3pt}
  \caption{
    Repair ablation on a representative 3-scene RTX~4090 subset
    (\textit{bicycle}, \textit{truck}, and \textit{playroom}). No repair is faster
    but causes substantial quality loss, while both default selective repair and
    full repair match the baseline within numerical noise.
  }
  \label{tab:repair_ablation}
  \begin{tabular}{lccc}
    \toprule
    Variant & Avg.\ frame & Max $|\Delta\mathrm{PSNR}|$ & Max $|\Delta\mathrm{SSIM}|$ \\
            & (ms)        & (dB)                         &                       \\
    \midrule
    No repair      & 2.849 & 2.0346              & 0.0579 \\
    Default repair & 3.098 & $6{\times}10^{-6}$  & $0$ \\
    Full repair    & 3.257 & $2{\times}10^{-6}$  & $0$ \\
    \bottomrule
  \end{tabular}
\end{table}

\section{Mechanism analysis}
\label{sec:analysis}

TileGS No-GW runs faster despite equal or greater raster DRAM traffic,
indicating that the speedup arises from a change in raster traversal granularity
rather than reduced byte volume.
We verify this using Nsight Compute Speed-of-Light, launch, divergence, and
instruction-count counters over the full 9-scene RTX~4090 aggregate under the
locked-clock preset; only locked-clock profiling runs are included in the
aggregate.
Counter values are collected in a dedicated pass and used for mechanism analysis
only, with primary performance results remaining those in \S\ref{sec:results}.

\noindent\textbf{Terminology: effective raster traversal work.}
We use this term to mean the raster kernel's \emph{dynamic} work: the candidate
Gaussian--tile entries actually visited, tested, and composited for a pixel
before transmittance early termination.
This differs from the \emph{static} workload---the total Gaussian--tile entries
assigned to a tile---which TileGS does not change (\S\ref{sec:bin_construction}).
We use two proxies: total SASS thread instructions for the raster kernel
(Table~\ref{tab:mechanism_analysis}) and average Gaussian tests per pixel before
termination (Table~\ref{tab:traversal_work}).

Table~\ref{tab:mechanism_analysis} summarizes the evidence:
despite lower SM throughput, lower active-warp occupancy, and higher DRAM
traffic, TileGS executes fewer total SASS thread instructions and completes
faster.
Concretely, Compute (SM) throughput decreases from 76.94\% to
59.61\%, achieved active-warp occupancy decreases from
91.32\% to 62.18\%, and DRAM traffic increases from
136.65\,MB to 185.53\,MB, while SASS thread instructions fall from 21.25B to
16.90B, a 1.26$\times$ reduction.

The counters rule out several simpler explanations.
TileGS does not improve occupancy, SM throughput, or memory volume.
It also uses more per-thread and per-block resources: register count increases
from 40 to 56 registers per thread, and dynamic shared memory
increases from 7{,}168\,B to 12{,}288\,B per block.
Divergence-related counters remain nearly unchanged: branch divergent threads
increase only slightly from 5.54M to 5.70M, and thread
instructions per warp instruction remain similar
(78.17\% vs.\ 77.75\%).
Thus, the measured speedup is not explained by improved coalescing, increased
occupancy, higher SM throughput, reduced memory traffic, or directly reduced
measured warp divergence.

The evidence instead supports a reduced-effective-traversal-work explanation.
TileGS reorganizes each long globally sorted tile-depth range into shorter
bin-local ranges, and the raster kernel performs less total thread-level
instruction work under this organization.
This is unlikely to be explained by compiler scheduling alone: both paths use
the same splat evaluation, early-exit test, and compositing equations, while the
binned path adds bin-boundary logic rather than removing inner-loop math.
The best-supported mechanism is fewer effective inner-loop iterations:
shorter, depth-ordered bin ranges may let pixels reach the transmittance
early-exit threshold with less wasted traversal.

As a direct diagnostic for this interpretation, we instrument the raster loop to
count the average number of Gaussian tests executed per pixel before
termination, measured over three repeated profiling passes on the fixed first
24 test views at $1600\times900$ resolution.
Table~\ref{tab:traversal_work} reports the results on three representative
scenes: two raster-heavy outdoor scenes (\textit{bicycle}, \textit{garden}) and
one lighter indoor scene (\textit{kitchen}), to demonstrate that the reduction
holds beyond the raster-heaviest regime.
TileGS reduces this direct traversal-work measure by 5.12\% on
\textit{bicycle}, 6.20\% on \textit{garden}, and 4.77\% on \textit{kitchen};
the measured test-count reductions are deterministic across the repeated passes
under the fixed-view protocol. 
This diagnostic does not isolate the exact per-pixel termination depth, but it
confirms that the binned traversal executes fewer Gaussian tests per pixel on
representative scenes.
The 4.77--6.20\% reduction in this direct test-count metric does not by itself
account for the full raster-kernel speedup: it is the narrow proxy defined
above, whereas SASS instructions capture per-visit compute, batch staging,
synchronization, active-mask behavior, and tile-level exits after all pixels
terminate.
The test-count metric also averages over all pixels and views, diluting larger
effects in hotspot tiles.
Thus, the direct test-count reduction and the 1.26$\times$ SASS instruction
reduction should be read as complementary views of the same
effective-raster-traversal-work reduction at different levels of the execution
hierarchy.

\begin{table}[H]
  \centering
  \caption{
    Direct traversal-work diagnostic on representative RTX~4090 scenes.
    Values report the average number of Gaussian tests executed per pixel before
    termination over three repeated profiling passes on the fixed first 24 test
    views at $1600\times900$.
    A Gaussian test counts one raster-loop visit to a candidate Gaussian for a
    pixel, before any per-candidate opacity/compositing update; the average is
    computed over pixels before the pixel-level early-exit terminates traversal.
  }
  \label{tab:traversal_work}
  \setlength{\tabcolsep}{4pt}
  \begin{tabular}{lccc}
    \toprule
    Scene & Baseline & TileGS & Reduction \\
    \midrule
    \textit{bicycle} & 570.646 & 541.444 & 1.054$\times$ \; (5.12\%) \\
    \textit{garden}  & 571.206 & 535.792 & 1.066$\times$ \; (6.20\%) \\
    \textit{kitchen} & 473.578 & 450.994 & 1.050$\times$ \; (4.77\%) \\
    \bottomrule
  \end{tabular}
\end{table}

Additionally, the per-scene kernel speedup scales with raster workload:
raster-heavy scenes (\textit{bicycle}, \textit{garden},
raster $\approx$48\% of GPU time)
achieve the largest speedups (1.42--1.46$\times$), consistent with a
traversal-work reduction that scales with range length rather than a fixed
structural or compiler effect.

We use \textit{bicycle} only for the source-attributed memory decomposition
because it is representative of the heavier-workload regime where TileGS
achieves its strongest end-to-end gains.
The full-counter trends in Table~\ref{tab:mechanism_analysis} are aggregated
over all nine scenes, while the source-attributed breakdown in
Table~\ref{tab:dram_decomp} should be interpreted as representative rather than
as a per-scene aggregate.

\begin{table}[tbp]
  \centering
  \caption{
    Full 9-scene RTX~4090 locked-clock Nsight aggregate for the baseline
    and TileGS No-GW raster kernels. TileGS is faster despite lower SM throughput,
    lower active-warp occupancy, and higher DRAM traffic, while executing fewer
    SASS thread instructions. Duration is summed across scenes before taking the
    ratio, so the aggregate ratio is duration-weighted.
  }
  \label{tab:mechanism_analysis}
  \setlength{\tabcolsep}{2pt}
  \begin{tabular}{lrrl}
    \toprule
    Metric & Baseline & TileGS & Interpretation \\
    \midrule
    Duration (ms) & 1.299 & 0.878 & $1.48\times$ faster \\
    SM throughput (\%) & 76.94 & 59.61 & lower \\
    Active-warp occ. (\%) & 91.32 & 62.18 & lower \\
    DRAM bytes (MB) & 136.65 & 185.53 & $+35.8\%$ \\
    SASS thread inst. & 21.25B & 16.90B & $1.26\times$ fewer \\
    Branch div. threads & 5.54M & 5.70M & nearly unchanged \\
    Thread insts / warp inst (\%) & 78.17 & 77.75 & nearly unchanged \\
    \bottomrule
  \end{tabular}
\end{table}

\subsection{Pipeline stage breakdown}
\label{sec:analysis_stages}

To understand both the sources of speedup and the cost of TileGS's auxiliary
passes, we decompose GPU kernel time by pipeline stage using PyTorch profiler
traces on \textit{bicycle} and \textit{garden} (RTX~4090).
Table~\ref{tab:stage_breakdown} reports the results.

\begin{table}[tbp]
  \centering
  \caption{
    Main contributors to per-stage GPU time changes on
    \textit{bicycle} and \textit{garden} (RTX~4090, PyTorch profiler traces).
    ``Sort'' denotes the baseline's full CUB global radix sort and TileGS's
    lighter per-bin segmented sort;
    ``Bin construct.'' denotes TileGS's tile-local count/scatter and prefix-sum
    pass, which is absent from the baseline (the small non-zero baseline entry
    reflects gsplat's existing tile-intersection counting kernel, not TileGS bin
    construction).
    Near-zero stages (spherical harmonics, projection, depth-range estimation)
    are omitted for compactness.
  }
  \label{tab:stage_breakdown}
  \setlength{\tabcolsep}{3pt}
  \begin{tabular}{lrrrrrr}
    \toprule
    & \multicolumn{3}{c}{\textbf{bicycle}} &
      \multicolumn{3}{c}{\textbf{garden}} \\
    \cmidrule(lr){2-4} \cmidrule(lr){5-7}
    Stage & Base & TileGS & $\Delta$ & Base & TileGS & $\Delta$ \\
    \midrule
    Raster        & 2.584 & 1.781 & $-0.803$ & 2.152 & 1.463 & $-0.689$ \\
    Sort          & 0.773 & 0.332 & $-0.441$ & 0.687 & 0.048 & $-0.640$ \\
    Tile intersect& 0.387 & 0.351 & $-0.036$ & 0.367 & 0.344 & $-0.023$ \\
    Bin construct.& 0.057 & 0.495 & $+0.438$ & 0.042 & 0.446 & $+0.404$ \\
    Repair        & 0.000 & 0.405 & $+0.405$ & 0.000 & 0.332 & $+0.332$ \\
    Total GPU     & 5.379 & 5.072 & $-0.307$ & 4.776 & 4.306 & $-0.470$ \\
    \bottomrule
  \end{tabular}
\end{table}

Two sources of speedup emerge.
The primary source is the \textbf{raster kernel}: TileGS saves 0.803\,ms on
bicycle and 0.689\,ms on garden, consistent with the $\approx$1.44$\times$
kernel speedup reported in \S\ref{sec:results_kernel}.
A secondary speedup source in our implementation is the \textbf{global sort}:
TileGS replaces the baseline's full CUB radix sort (0.773\,ms / 0.687\,ms)
with a lighter combination of a per-tile bin scatter pass and a per-bin
segmented sort, saving 0.441\,ms on bicycle and 0.640\,ms on garden.

The two main added costs are bin construction (0.438/0.404\,ms) and repair
(0.405/0.332\,ms).
Tile intersection shows a small saving (0.036/0.023\,ms), while spherical
harmonics, projection, and depth-range estimation have near-zero deltas and
are omitted.
Reducing these auxiliary costs --- particularly by fusing small kernels in the
bin construction and repair passes --- is the highest-leverage direction for
closing the frame/kernel gap.
The 6.9--9.4\% end-to-end gain is achieved despite the added bin-construction
and repair stages, so reducing auxiliary-stage overhead is the most direct path
to larger frame-level gains.
The decomposed stages also show why the mechanism analysis focuses on the
raster kernel rather than the full pipeline.
On both scenes, the sort replacement (0.441/0.640\,ms) alone is smaller than
the combined added cost of bin construction and repair (0.843/0.736\,ms), so
on these two scenes the shown non-raster stages net to a cost rather than a
saving, making the raster-kernel saving the dominant contributor to the net
frame-level gain.
This is why \S\ref{sec:analysis} analyzes the raster kernel specifically when
explaining the source of the speedup.

\subsection{Remaining bottleneck: geometry-attribute scatter}
\label{sec:analysis_remaining}

The RTX~4090 and RTX~1000~Ada results indicate that TileGS No-GW primarily
acts as a traversal-granularity optimization rather than a bandwidth-reduction
method. On RTX~4090, TileGS is faster despite moving more raster data than the
baseline.
On RTX~1000~Ada, TileGS also achieves clear end-to-end gains under a more
bandwidth-constrained operating point, although full-resolution Nsight captures
are unavailable for four raster-heavy Ada scenes.
The captured five-scene Ada kernel subset nevertheless yields a mean
1.441$\times$ speedup, and the 50\% Gaussian diagnostic subset discussed in
\S\ref{sec:discussion} shows the same qualitative pattern as RTX~4090: fewer
SASS thread instructions despite higher DRAM traffic.
These cross-platform indicators are consistent with the same traversal-work
interpretation, but are not sufficient to attribute the Ada gain to the same
source-level traffic components measured on RTX~4090.

To understand the remaining bottleneck, we attribute memory pressure to
source-level regions in a representative \textit{bicycle} raster-kernel capture
on RTX~4090.
As shown in Table~\ref{tab:dram_decomp}, geometry attributes dominate both total
traffic and excess sectors: they account for 85.8\% of total source-attributed
raster memory pressure and 88.6\% of excess sectors.
We verified qualitatively that the geometry-attribute dominance is stable across
scene types: on lighter scenes, absolute traffic volumes are lower but the
relative share of geometry attributes remains above 80\% of total
source-attributed pressure, because the bin/index structures scale with the same
Gaussian count while geometry remains the dominant per-entry fetch cost.
This indicates that the unresolved bottleneck is not primarily tile-bin
metadata, but the scattered access pattern of conic parameters, projected means,
colors, and opacities during rasterization, regardless of scene weight.

We also explored lightweight bandwidth-oriented fixes, including in-kernel
gather-order sorting, FP16 color compression, sidecar materialization, and
pre-raster rejection heuristics.
None produced a net gain in the current CUDA/IMR implementation: the added
control, synchronization, decode, or helper-kernel cost outweighed the reduction
in attribute traffic.
The No-GW path achieves better results partly because it keeps the raster inner
loop simple.
Thus, the next meaningful optimization direction is structural geometry-layout
reorganization, such as raster-order-aware storage or AoSoA-style layouts, that
better aligns attribute layout with tile-local traversal.

\begin{table}[H]
  \centering
  \caption{
    Representative source-attributed DRAM pressure
    (L2 Theoretical Sectors, Global) for the TileGS No-GW rasterization kernel on
    \textit{bicycle} on RTX~4090.
    ``Excess'' sectors denote uncoalesced cache-line bytes loaded but not used.
  }
  \label{tab:dram_decomp}
  \begin{tabular}{lrrrr}
    \toprule
    Region & Total (M) & \% & Excess (M) & \% \\
    \midrule
    Geometry attributes & 45.7 & 85.8 & 39.9 & 88.6 \\
    Bin/index structures &  5.4 & 10.2 &  4.1 &  9.0 \\
    Framebuffer         &  2.1 &  4.0 &  1.1 &  2.4 \\
    \midrule
    \textbf{Total}      & \textbf{53.2} & \textbf{100} & \textbf{45.1} & \textbf{100} \\
    \bottomrule
  \end{tabular}
\end{table}

\section{Discussion, limitations, and future work}
\label{sec:discussion}

\noindent\textbf{Scope and complementarity.}
TileGS should be understood primarily as a tile-local traversal organization
improvement for Gaussian rasterization. It is not merely a CUDA implementation
variant of 3DGS; it exposes tile-local traversal granularity as a separate
design axis for Gaussian rasterization.
Its contribution is an execution-structure insight---that standard 3DGS is
tile-based in ownership but not tile-local in traversal---and a demonstration
that changing traversal granularity alone, without altering the representation
or compositing rule, produces measurable speedups even when memory coalescing
and occupancy counters do not improve.

\noindent\textbf{GPU architecture scope.}
Both evaluation platforms are NVIDIA Ada-generation GPUs; we do not evaluate
other NVIDIA generations, non-NVIDIA GPUs, or mobile/TBDR architectures.
Because the magnitude of tile-bin-major reordering depends on hardware
scheduling, cache behavior, and memory service, we make no quantitative claim
beyond the two Ada platforms evaluated here.

\noindent\textbf{Profiling limitations on RTX~1000~Ada.}
Full-resolution Nsight counter capture failed on four raster-heavy scenes
(\textit{bicycle}, \textit{garden}, \textit{stump}, \textit{drjohnson}) due to
driver/profiler resource acquisition errors, while the application runs
completed normally.
The reported RTX~1000~Ada kernel mean of 1.441$\times$ is therefore a
five-scene subset, not a full-suite average.
A 50\% Gaussian diagnostic subset completed on all nine scenes and showed the
same qualitative trend as RTX~4090: TileGS reduced kernel duration by
1.45$\times$ and SASS thread instructions by 1.13$\times$
while increasing DRAM traffic.
We use this only as supporting diagnostic evidence, not as a main aggregate.

\noindent\textbf{Scene coverage.}
The benchmark is limited to natural real-world captures from Mip-NeRF~360,
Tanks and Temples, and Deep Blending.
Synthetic scenes with unusually sparse, unusually dense, or strongly bimodal
near/far depth distributions may interact differently with log-depth binning
and the repair policy.
The supplemental material includes a single-view synthetic scale-up diagnostic
on a high-workload \textit{bicycle} view; this test is intended only to
characterize sort/reordering and raster-stage behavior under increased
emitted-entry pressure, not to replace the full 9-scene benchmark.
In that diagnostic, TileGS keeps both stages consistently cheaper than the
baseline, although total-frame speedup remains limited by non-raster and
auxiliary overheads, consistent with the frame/kernel gap analysis in
\S\ref{sec:analysis_stages}.
A systematic stress test over synthetic sparse, dense, and bimodal depth
distributions would be useful for characterizing when binning overhead dominates
and when the repair budget becomes active; we treat this as future work.

\noindent\textbf{Resolution and view-count scope.}
The reported configuration---$K=64$, the depth-range percentiles in
\S\ref{sec:depth_range}, and the repair thresholds in
Table~\ref{tab:repair_rules}---was evaluated at each scene's native resolution
and standard held-out view split. We do not independently sweep resolution or
view count. Higher resolutions may change the number of tiles, per-tile
Gaussian counts, and the best bin/repair settings, so we leave a systematic
resolution sweep to future work.

\noindent\textbf{Backward-pass integration.}
The current TileGS implementation targets the forward rasterization pass only.
Extending it to training would require the backward pass to traverse Gaussian–tile
entries in the same tile-bin-major, repaired order used in the forward pass, since
the alpha-compositing gradient depends on that exact ordering.
We leave this integration to future work.

\noindent\textbf{Implementation details.}
Although the main text specifies the repair buckets, default thresholds, and
global repair-entry budget, full implementation-level pseudocode for bin
construction, repair selection, and binned raster traversal is provided in the
supplemental material.
Aggressive and conservative repair-threshold variants changed full-suite
RTX~4090 frame time by less than 1\% while remaining within the same quality
thresholds.

\noindent\textbf{Dominant remaining bottleneck and future direction.}
The current No-GW implementation does not consistently reduce DRAM byte volume;
the gain is a traversal-granularity effect, as supported in \S\ref{sec:analysis}.
Source-attributed profiling shows that geometry-attribute access dominates the
remaining memory pressure (Table~\ref{tab:dram_decomp}).
The next meaningful optimization direction is therefore structural
geometry-layout reorganization, such as raster-order-aware storage or AoSoA-style
layouts, as discussed in \S\ref{sec:analysis_remaining}.
On bandwidth-constrained or TBDR-style GPUs, this structural alignment may be
important for realizing stronger bandwidth-reduction benefits from tile-local
Gaussian rendering.

\section{Conclusion}
\label{sec:conclusion}

We presented TileGS, a tile-local depth-binning pipeline for Gaussian splatting
rasterization.
TileGS replaces the baseline globally sorted tile-depth stream traversal with a
bin-organized tile-local raster path and a selective exact-repair mechanism.
The method keeps the Gaussian representation, projected inputs, and compositing
semantics unchanged, but changes the granularity at which the raster kernel
consumes the Gaussian workload.

Across a 9-scene benchmark on desktop- and laptop-class Ada GPUs, TileGS
preserves image quality up to numerical noise while improving both
rasterization-kernel time and end-to-end frame time over gsplat.
The gains are achieved despite added depth-range estimation, bin construction,
and repair stages, making auxiliary-stage fusion the most direct path to larger
frame-level speedups.

The broader lesson is that Gaussian rasterization benefits from execution-order
redesign even when the representation and inner-loop math are unchanged.
Full-suite RTX~4090 profiling shows that TileGS is faster despite higher DRAM
traffic and lower occupancy/SM-throughput counters, while SASS thread
instructions fall by 1.26$\times$.
Geometry-attribute access remains the dominant unresolved bottleneck, suggesting
that structural alignment between attribute storage and tile-local traversal is
a promising direction for bandwidth-constrained or TBDR-oriented Gaussian
renderers.


\bibliographystyle{eg-alpha-doi}
\bibliography{tilegs}

\end{document}


\maketitle

\noindent\textbf{Overview.}
This supplemental material provides implementation-level pseudocode for the
TileGS forward path used in the main paper. The goal is to clarify the execution
structure: robust depth-range estimation, tile-local bin construction,
selective exact repair, and binned raster traversal. The pseudocode follows the
current No-GW implementation at the level of data dependencies and routing
decisions, while omitting low-level CUDA launch details, workspace allocation,
and optional profiling/debug paths.

\section{Notation and Default Configuration}
Let $E$ denote the emitted Gaussian--tile intersection entries. Each entry stores
a tile id $t(e)$, a Gaussian id $g(e)$, and an original baseline sort key $s(e)$
encoding the baseline depth order inside the tile. TileGS assigns each entry to a
coarse bin $b(e) \in \{0,\ldots,K-1\}$ and constructs a flattened tile-bin-major
stream. For each tile $t$, the offset array $\mathrm{binOffsets}[t,0\ldots K]$
defines the concatenated interval for all bins in that tile; the segment for bin
$b$ is $[\mathrm{binOffsets}[t,b],\mathrm{binOffsets}[t,b+1])$.

\begin{table}[t]
  \centering
  \caption{Default No-GW configuration represented by this supplement.}
  \label{tab:supp_defaults}
  \setlength{\tabcolsep}{3pt}
  \begin{tabular}{@{}p{0.30\linewidth}p{0.62\linewidth}@{}}
    \toprule
    Setting & Value \\
    \midrule
    Raster path & No-GW: no pre-raster geometry materialization \\
    Depth bins & $K=64$ \\
    Depth-to-bin mapping &
    Log-space (default); linear variant available but unused \\
    Depth-range percentiles &
    $p_{\mathrm{lo}}=1.0$, $p_{\mathrm{hi}}=99.0$ \\
    Depth-range padding &
    $f_{\mathrm{pad}}=0.05$ (5\% of clipped span) \\
    Depth-range sample count &
    Up to $S=8192$ candidates \\
    Depth-range selection method &
    Exact order statistics (\texttt{kthvalue}); histogram alternative exists but unused \\
    Near-depth shift &
    Enabled by default, $\alpha=1.0$, $\gamma=0.25$ \\
    Count/scatter path & Direct atomic tile-bin construction \\
    Warp-key grouping & Disabled in the reported default \\
    Geometry sidecar & Disabled for No-GW; optional for the Packed-GW ablation \\
    Flat-id sidecar & Optional; otherwise the raster kernel reads the flat id from the pair stream \\
    Raster replay & Concatenated tile-bin stream in front-to-back bin order \\
    Color staging & Enabled by default in the binned raster kernel \\
    Conservative alpha cull & Enabled by default in the binned raster kernel \\
    Repair & Selective exact repair for selected tile/bin segments \\
    \bottomrule
  \end{tabular}
\end{table}

\begin{algorithm}[tbp]
  \caption{Overall TileGS forward path}
  \label{alg:overall_tilegs}
  \begin{algorithmic}[1]
    \Require Projected Gaussian attributes; baseline Gaussian--tile entries $E$;
      image and tile dimensions; $K=64$
    \Ensure Rendered image matching baseline compositing up to numerical noise
    \State Estimate robust depth range $[z_{\min},z_{\max}]$ using
      Algorithm~\ref{alg:zrange}
    \State Build a tile-bin-major pair stream and bin offsets using
      Algorithm~\ref{alg:bin_construct}
    \State Apply selected exact repair by sorting selected tile/bin segments using
      Algorithm~\ref{alg:repair_select}
    \State Rasterize the repaired tile-bin-major stream using the binned traversal
      procedure in Section~\ref{sec:binned_raster}
    \State \Return rendered image
  \end{algorithmic}
\end{algorithm}

\section{Tile-Local Bin Construction}
The current implementation builds a pair stream. Each pair stores the original
baseline ordering key and the Gaussian flat id. The final pair stream is
tile-bin-major: all entries of tile $t$, bin $0$ appear before entries of tile
$t$, bin $1$, and so on. In the No-GW path, the raster kernel normally fetches
geometry attributes from the original arrays using the flat id. Packed-GW
materializes optional geometry sidecars, but that path is not the default
reported configuration.

\begin{algorithm}[H]
  \caption{Robust depth-range estimation}
  \label{alg:zrange}
  \begin{algorithmic}[1]
    \Require Projected depths; Gaussian scales; optional emitted flat ids;
      near/far planes; sample count $S=8192$; near-depth shift parameters
      $\alpha=1.0$, $\gamma=0.25$; percentiles
      $p_{\mathrm{lo}}=1.0$, $p_{\mathrm{hi}}=99.0$; padding fraction
      $f_{\mathrm{pad}}=0.05$; numerical floor $\epsilon=10^{-4}$
    \Ensure Robust clipped range $[z_{\min},z_{\max}]$
    \State Sample up to $S$ depth candidates from visible Gaussians or emitted flat ids
    \If{near-depth mode is enabled (default)}
      \State Convert each sampled depth $z_c$ to
        $z_{\mathrm{key}} \gets z_c - \min(\alpha\cdot s_{\max},\,\gamma\cdot z_c)$,
        where $s_{\max}$ is the Gaussian's maximum projected scale
    \EndIf
    \State Discard candidates with $z_{\mathrm{key}}\le 0$; clamp remaining values using the near/far planes and $\epsilon$
    \State Select the exact $p_{\mathrm{lo}}/p_{\mathrm{hi}}$ percentile values of the sampled candidates via order statistics to obtain $[z_{\mathrm{lo}},z_{\mathrm{hi}}]$
    \State $\mathrm{span} \gets \max(z_{\mathrm{hi}}-z_{\mathrm{lo}},10^{-6})$
    \State $z_{\min} \gets \max(z_{\mathrm{near}},\, z_{\mathrm{lo}} - f_{\mathrm{pad}}\cdot \mathrm{span})$,\quad
      $z_{\max} \gets \min(z_{\mathrm{far}},\, z_{\mathrm{hi}} + f_{\mathrm{pad}}\cdot \mathrm{span})$
    \State \Return $[z_{\min},z_{\max}]$
  \end{algorithmic}
\end{algorithm}

\noindent
Percentile selection uses exact order statistics (\texttt{kthvalue}) over the
sampled candidates rather than a histogram approximation; this is the method
that produced all reported results. A histogram-based percentile estimator
(log-spaced, 128 bins by default) is available as an alternative but is not
the default and was not used for the reported numbers.
The near-depth shift is used only to form the binning key; it does not change
the projected Gaussian attributes or the alpha-compositing equations used by
the raster kernel.

\begin{algorithm}[H]
  \caption{Tile-local bin construction}
  \label{alg:bin_construct}
  \begin{algorithmic}[1]
    \Require Baseline entries $E$ with tile id $t(e)$, flat id $g(e)$, baseline
      sort key $s(e)$, and depth $z(e)$; clipped range $[z_{\min},z_{\max}]$;
      number of bins $K$
    \Ensure Tile offsets, bin offsets, tile-bin-major pair stream $P$, and
      optional sidecars
    \State Initialize $\mathrm{count}[t,b] \gets 0$ for all tiles and bins
    \ForAll{entries $e \in E$ in parallel}
      \State $b(e) \gets \mathrm{log\text{-}depth\text{-}bin}(z(e),
        z_{\min},z_{\max},K)$
      \State Atomically increment $\mathrm{count}[t(e),b(e)]$
    \EndFor
    \State Exclusive-scan the count array to obtain $\mathrm{binOffsets}[t,b]$
    \State Initialize write cursors from $\mathrm{binOffsets}$
    \ForAll{entries $e \in E$ in parallel}
      \State $b \gets b(e)$
      \State $p \gets$ atomic increment of $\mathrm{cursor}[t(e),b]$
      \State $P[p] \gets (s(e),g(e))$
      \If{flat-id sidecar is enabled}
        \State $\mathrm{flatIds}[p] \gets g(e)$
      \EndIf
      \If{Packed-GW geometry sidecar is enabled}
        \State Materialize selected geometry attributes at index $p$
      \EndIf
    \EndFor
    \State \Return tile offsets, $\mathrm{binOffsets}$, $P$, and optional sidecars
  \end{algorithmic}
\end{algorithm}

\noindent
The mapping $\mathrm{log\text{-}depth\text{-}bin}(z,z_{\min},z_{\max},K)$ used
above is, by default, a clamped log-space bin index:
\[
u = \mathrm{clamp}\!\left(
  \frac{\ln(\max(z,\epsilon)) - \ln(z_{\min}')}
       {\ln(z_{\max}') - \ln(z_{\min}')},
  0,\ 1-10^{-6}\right),
\]
\[
b = \left\lfloor K u \right\rfloor, \qquad
b \gets \mathrm{clamp}(b,\ 0,\ K-1).
\]
where $z_{\min}' = \max(z_{\min},\epsilon)$,
$z_{\max}' = \max(z_{\max}, z_{\min}'+10^{-6})$, and $\epsilon = 10^{-4}$.
The per-entry depth $z(e)$ passed to this mapping is the same near-depth-shifted
key $z_{\mathrm{key}}$ used during range estimation (Algorithm~\ref{alg:zrange}),
so bin assignment and range estimation apply an identical depth transform.
A linear (non-log) variant with the same clamp-and-floor structure is
available but is not the default.

\begin{table}[tbp]
  \centering
  \caption{
    Packed-GW diagnostic on RTX~4090. The April 22 profiling sweep shows that
    Packed-GW can reduce raster DRAM traffic, but the April 23 timing-only rerun
    is uniformly slower. We therefore use No-GW as the default and treat
    Packed-GW as a bandwidth-oriented diagnostic.
  }
  \label{tab:packed_gw_diag}
  \setlength{\tabcolsep}{3pt}
  \begin{tabular}{@{}p{0.28\linewidth}p{0.33\linewidth}p{0.29\linewidth}@{}}
    \toprule
    Evidence & 
    Result & 
    Interpretation \\
    \midrule
    April 22 profiling sweep & 
    DRAM reduced in 7/9 scenes & 
    bandwidth benefit exists \\
    Mean DRAM change & 
    $-28.64$ MB & 
    traffic-side improvement \\
    Largest reductions & 
    garden $-92.46$ MB; bicycle $-82.34$ MB & 
    strongest on heavy scenes \\
    April 23 timing rerun & 
    slower in 9/9 scenes & 
    not selected as default \\
    Mean timing change & 
    $+0.208$ ms / $+7.87\%$ & 
    materialization cost dominates \\
    \bottomrule
  \end{tabular}
\end{table}

\begin{table}[tbp]
  \centering
  \caption{
    Count/scatter construction variants on RTX~4090
    (\texttt{kernel\_variant\_full9\_rtx4090\_20260504\_v3}).
    All variants produce identical bin-offset arrays and Gaussian-ID streams;
    only construction/runtime overhead changes.
  }
  \label{tab:count_scatter_variants}
  \setlength{\tabcolsep}{4pt}
  \begin{tabular}{lcc}
    \toprule
    Variant & 
    Mean time & 
    Relative to default \\
            & 
            (ms)      &                     \\
    \midrule
    Warp-key grouping       & 
    2.85796 & 
    $0.996\times$ \\
    Two-stage aggregation   & 
    2.85863 & 
    $0.997\times$ \\
    Default direct atomics  & 
    2.86811 & 
    $1.000\times$ \\
    Block-local hash agg.   & 
    2.92911 & 
    $1.021\times$ \\
    Warp-private histogram  & 
    3.32800 & 
    $1.160\times$ \\
    \bottomrule
  \end{tabular}
\end{table}

Warp-private histograms were slower than the default on all nine scenes, with
the largest slowdowns on \textit{bicycle} (+0.932 ms) and \textit{garden}
(+0.840 ms).
Although warp-key grouping and two-stage aggregation were within 0.4\% of the
default in this isolated sweep, a later paired rerun favored the final
warp-key-disabled path by 0.01982 ms in median time and showed a tighter run
range (0.01952 ms vs. 0.04773 ms).
We therefore use direct atomics with warp-key grouping disabled as the reported
default.

\section{Selective Exact Repair}
The implementation represents every coarse tile/bin slice as a logical segment.
Repair is applied before rasterization by sorting selected segments according to
their original baseline key. Non-selected segments remain in the tile-bin-major
stream produced by the construction pass. The reported selective policy uses
fixed global thresholds; they are not tuned per scene.

\begin{table}[H]
  \centering
  \caption{Default selective-repair thresholds represented in the pseudocode.}
  \label{tab:supp_repair_thresholds}
  \setlength{\tabcolsep}{3pt}
  \begin{tabular}{ll}
    \toprule
    Parameter & Default value \\
    \midrule
    No-repair threshold & $\ell \le 1$ \\
    Local exact-repair maximum & 512 entries \\
    Always-keep long segment cutoff & $\ell \ge 320$ \\
    Global-tail bucket & $\ell > 512$ \\
    Tile-share cutoff & $\ell/\ell_{\mathrm{tile}} \ge 45\%$ \\
    Local-256 share cutoff & $129 \le \ell \le 256$ and $\ell/\ell_{\mathrm{tile}} \ge 10\%$ \\
    Very-front cutoff & $b < 2$ and $\ell \ge 16$ \\
    Repair-entry budget & $B_{\mathrm{rep}}=\lfloor 0.25M \rfloor$, where $M=|E|$ \\
    \bottomrule
  \end{tabular}
\end{table}

\begin{algorithm}[H]
  \caption{Selective repair selection and local exact sorting}
  \label{alg:repair_select}
  \begin{algorithmic}[1]
    \Require Pair stream $P$; bin offsets; $K$; emitted-entry count $M$;
      repair-entry budget $B_{\mathrm{rep}}=\lfloor 0.25M \rfloor$
    \Ensure Pair stream with selected segments restored to baseline order
    \State $R \gets \emptyset$,\quad $\mathrm{used} \gets 0$
    \For{logical segment ids $q=tK+b$ in tile-bin order}
      \State $\mathrm{begin} \gets \mathrm{binOffsets}[t,b]$
      \State $\mathrm{end} \gets \mathrm{binOffsets}[t,b+1]$
      \State $\ell \gets \mathrm{end}-\mathrm{begin}$
      \If{$\ell \le 1$}
        \State \textbf{continue}
      \EndIf
      \State $\ell_{\mathrm{tile}} \gets \mathrm{binOffsets}[t,K]-\mathrm{binOffsets}[t,0]$
      \State $\mathrm{keep} \gets \mathrm{false}$
      \If{$\ell \ge 320$}
        \State $\mathrm{keep} \gets \mathrm{true}$
      \EndIf
      \If{$\ell > 512$}
        \State $\mathrm{keep} \gets \mathrm{true}$
      \EndIf
      \If{$\ell/\ell_{\mathrm{tile}} \ge 45\%$}
        \State $\mathrm{keep} \gets \mathrm{true}$
      \EndIf
      \If{$129 \le \ell \le 256$ \textbf{and} $\ell/\ell_{\mathrm{tile}} \ge 10\%$}
        \State $\mathrm{keep} \gets \mathrm{true}$
      \EndIf
      \If{$b < 2$ \textbf{and} $\ell \ge 16$}
        \State $\mathrm{keep} \gets \mathrm{true}$
      \EndIf
      \If{$\mathrm{keep}$ \textbf{and} $\mathrm{used}+\ell \le B_{\mathrm{rep}}$}
        \State Add $(t,b,\mathrm{begin},\mathrm{end})$ to $R$
        \State $\mathrm{used} \gets \mathrm{used}+\ell$
      \EndIf
    \EndFor
    \ForAll{segments $(t,b,\mathrm{begin},\mathrm{end}) \in R$}
      \State Sort $P[\mathrm{begin}:\mathrm{end})$ by original baseline key $s(e)$
      \State Keep flat-id/geometry sidecars aligned with the sorted pair order
    \EndFor
    \State \Return repaired pair stream and aligned sidecars
  \end{algorithmic}
\end{algorithm}

\section{Binned Raster Traversal}
\label{sec:binned_raster}

\noindent\textbf{Relationship to the baseline traversal.}
At the control-flow level, Algorithm~\ref{alg:binned_raster_traversal} resembles
the baseline: each tile's CUDA block linearly scans one contiguous interval of
the pair stream, in batches, with the same per-pixel early-termination test.
What differs from the baseline is the \emph{contents} and organization of the
interval being scanned.
Concretely, the interval $[a,b) = [\mathrm{binOffsets}[t,0],\mathrm{binOffsets}[t,K])$
that each tile's block replays is not a single globally depth-sorted range as
in the baseline, but the concatenation of the tile's $K$ coarse bins produced
by tile-local bin construction (Algorithm~\ref{alg:bin_construct}), laid out
back to back in front-to-back bin order --- the tile-bin-major layout defined
in the Tile-local bin construction subsection of the main paper's Method
section. Within a given bin's sub-range, entries sit in
whatever order the atomic scatter of Algorithm~\ref{alg:bin_construct}
happened to produce; they are \emph{not} depth-sorted, except where selective
exact repair (Algorithm~\ref{alg:repair_select}) has restored exact baseline
depth order to that specific segment.
Algorithm~\ref{alg:binned_raster_traversal} therefore requires no per-bin
branching: the tile-local reordering is already baked into the stream before
the raster kernel runs, so a plain linear scan over $[a,b)$ is sufficient to
consume the bins in front-to-back order. The main paper's Mechanism analysis
section discusses why this reordering, rather than any change to the traversal
loop itself, is the source of the measured speedup.

\noindent\textbf{Implementation details.}
In the default No-GW path, the kernel loads Gaussian attributes from the
original arrays using either a flat-id sidecar or the flat id stored in the
pair stream, using one CUDA block per tile. A profiling mode can instead
replay a specified bin window in isolation, but this is not the default
reported path. Thus, the default raster loop does not use intermediate bin
boundaries as separate raster-loop exits; the boundaries determine the
materialized stream layout and repair segments before rasterization.
Algorithm~\ref{alg:binned_raster_traversal} gives pseudocode
for the binned raster traversal after repair.

\begin{algorithm}[tbp]
  \caption{Binned raster traversal after repair (default No-GW path)}
  \label{alg:binned_raster_traversal}
  \begin{algorithmic}[1]
    \Require Repaired pair stream $P$; bin offsets; optional flat-id sidecar;
      Gaussian attributes; image/tile dimensions
    \Ensure Rendered colors, alphas, and last contributing ids
    \ForAll{tiles $t$ in parallel}
      \State Determine tile pixel coordinates; skip masked tiles
      \ForAll{pixels $p$ in tile $t$}
        \State $C_p \gets 0$;\quad $T_p \gets 1$
      \EndFor
      \State $a \gets \textit{binOffsets}[t,0]$;\quad $b \gets \textit{binOffsets}[t,K]$
      \For{each batch in $[a,b)$}
        \If{all pixels in tile have terminated} \textbf{break} \EndIf
        \State Load candidate Gaussian ids into shared memory
        \If{tile-alpha culling is enabled}
          \State Remove candidates below the tile-level contribution threshold
        \EndIf
        \If{color staging is enabled}
          \State Stage candidate colors cooperatively in shared memory
        \EndIf
        \ForAll{Gaussians $g$ in surviving batch order}
          \ForAll{active pixels $p$ in tile $t$}
            \State Count candidate visit if traversal profiling is enabled
            \State Evaluate the same opacity/conic test as the baseline
            \State Update $C_p$ and $T_p$ using baseline alpha compositing
            \If{$T_p$ reaches the early-termination threshold}
              \State Mark pixel $p$ inactive
            \EndIf
          \EndFor
        \EndFor
      \EndFor
      \State Write final color, alpha, and last contributing id for each pixel
    \EndFor
    \State \Return rendered image
  \end{algorithmic}
\end{algorithm}

\FloatBarrier
\section{Synthetic Scale-Up Stress Diagnostic}
To complement the full 9-scene benchmark in the main paper, we include a
single-view synthetic scale-up diagnostic on a fixed high-workload view from the
real \textit{bicycle} scene (\texttt{\_DSC8687.JPG}). This diagnostic is not a
replacement for the full benchmark; instead, it isolates how the sort/reordering
and raster stages behave as the emitted Gaussian workload increases. We
synthetically scale the workload from 5M to 30M Gaussians and measure
median-of-5 timings under the same locked RTX~4090 timing-only protocol used
for the main evaluation.

\begin{figure}[tbp]
  \centering
  \includegraphics[width=0.96\linewidth]{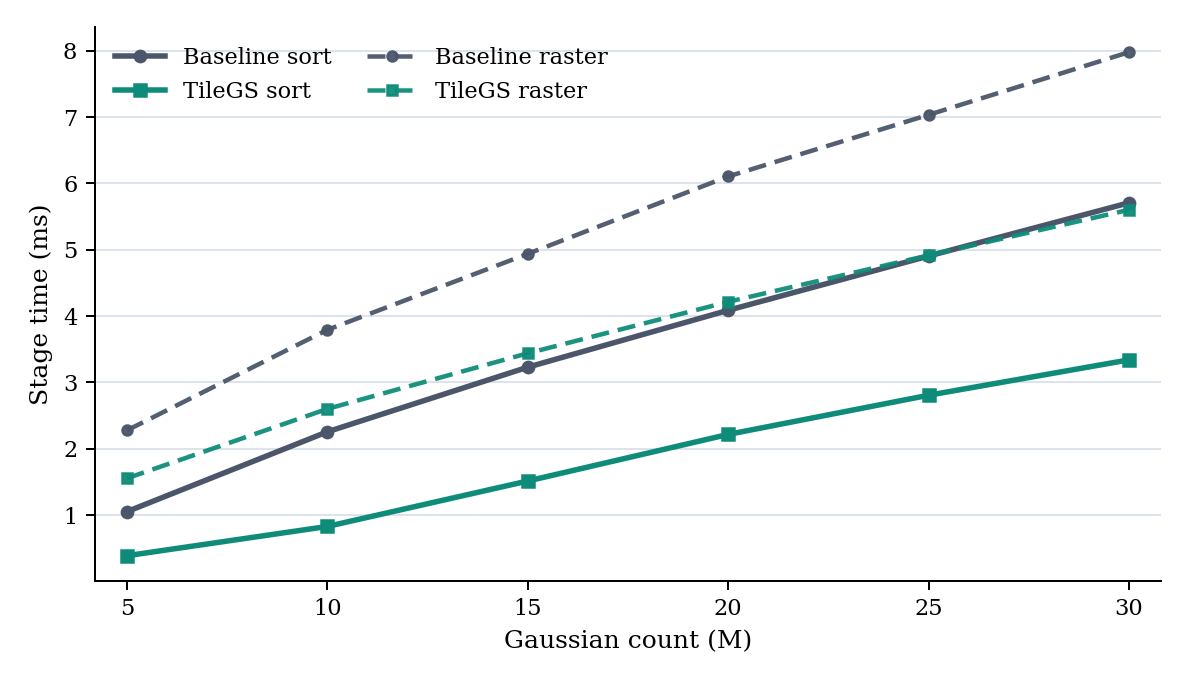}
  \caption{
    Synthetic scale-up diagnostic on a fixed high-workload \textit{bicycle} view
    (\texttt{\_DSC8687.JPG}). The workload increases from 5M to 30M Gaussians.
    TileGS keeps both sort/reordering and raster stages cheaper than the gsplat
    baseline; this should be read as supplementary scale-up evidence rather than
    as a new headline benchmark.
  }
  \label{fig:supp_scalability}
\end{figure}

Figure~\ref{fig:supp_scalability} and Table~\ref{tab:supp_scalability} show
that both the sort/reordering stage and the raster stage remain consistently
cheaper for TileGS across the tested range, and the absolute timing gaps widen
with scale.
At 30M emitted entries, the sort stage improves by 1.71$\times$ and the
raster stage by 1.42$\times$, consistent with the main paper's interpretation
that tile-local depth-binned traversal reduces forward-pass cost as raster
workload intensity increases.
The total-frame speedup is smaller and not monotonic because the synthetic
scale-up includes non-raster and auxiliary overheads that are not eliminated by
the binned raster path; this behavior is consistent with the frame/kernel gap
analysis in the main paper.

\begin{table}[t]
  \centering
  \setlength{\tabcolsep}{3pt}
  \small
  \caption{
    Synthetic scale-up timing on the fixed \textit{bicycle}
    view used in Figure~\ref{fig:supp_scalability}. Times are median-of-5
    measurements on RTX~4090 under the locked timing-only protocol.
  }
  \label{tab:supp_scalability}
  \begin{tabular}{rccccccccc}
    \toprule
    Points & \multicolumn{3}{c}{Sort (ms)} &
             \multicolumn{3}{c}{Raster (ms)} &
             \multicolumn{3}{c}{Total (ms)} \\
    \cmidrule(lr){2-4}\cmidrule(lr){5-7}\cmidrule(lr){8-10}
    (M) & Base & TileGS & $\times$ & Base & TileGS & $\times$ & Base & TileGS & $\times$ \\
    \midrule
     5 & 1.051 & 0.387 & 2.72 & 2.277 & 1.554 & 1.46 &  4.843 &  4.345 & 1.11 \\
    10 & 2.255 & 0.828 & 2.72 & 3.793 & 2.601 & 1.46 &  8.797 &  7.650 & 1.15 \\
    15 & 3.230 & 1.513 & 2.13 & 4.939 & 3.440 & 1.44 & 12.174 & 10.772 & 1.13 \\
    20 & 4.087 & 2.216 & 1.84 & 6.106 & 4.216 & 1.45 & 15.290 & 13.923 & 1.10 \\
    25 & 4.906 & 2.807 & 1.75 & 7.035 & 4.916 & 1.43 & 18.311 & 16.733 & 1.09 \\
    30 & 5.709 & 3.340 & 1.71 & 7.980 & 5.603 & 1.42 & 21.106 & 19.441 & 1.09 \\
    \bottomrule
  \end{tabular}
\end{table}

\FloatBarrier
\section{Reproducibility Checklist}
\label{sec:supp_reproducibility}
For the main reported timing runs, use locked GPU clock presets, CUDA-event GPU
timing, resident Gaussian parameters, and the same projected Gaussian inputs for
baseline and TileGS. Nsight Compute counter collection should be performed in a
separate profiling pass and should not be mixed with the primary timing-only
measurements. Unless otherwise stated, use the No-GW path, $K=64$,
direct-atomic tile-bin construction with warp-key grouping disabled, color
staging and conservative alpha culling enabled, log-space depth binning with
the depth-range and near-depth-shift constants in
Table~\ref{tab:supp_defaults}, and selective exact repair with the thresholds
in Table~\ref{tab:supp_repair_thresholds}.

%